\documentclass[aps,preprint,floatfix]{revtex4}
\usepackage{epsfig}

\newcommand{\be}{\begin{equation}}
\newcommand{\ee}{\end{equation}}

\begin{document}

\title{Chiral symmetry breaking at large $N_c$}
\author{R. Narayanan}
\affiliation{
Department of Physics, Florida International University, Miami,
FL 33199\\{\tt rajamani.narayanan@fiu.edu}}
\author{ H. Neuberger}
\affiliation{
Rutgers University, Department of Physics
and Astronomy,
Piscataway, NJ 08855\\{\tt neuberg@physics.rutgers.edu}
}

\begin{abstract}
We present numerical evidence for the hypothesis that, in the planar limit,
four dimensional Euclidean Yang-Mills theory
on a finite symmetrical four-torus breaks chiral symmetry spontaneously
when the length of the sides $l$ is larger than a critical value $l_c$
with a bilinear condensate whose value is independent of $l$. 
Therefore spontaneous symmetry breaking occurs at finite volume and
infinite $N_c$ reduction holds for the chiral condensate.
\end{abstract}

\maketitle

\section {Introduction. }

The sum of pure gauge connected vacuum Feynman diagrams goes as
$N_c^2$ at fixed 't Hooft coupling, $\lambda=g^2 N_c$~\cite{thooft}. Fermions
make a contribution that goes only as $N_c$. Similar behavior occurs in the 
strong coupling expansion on the lattice. In short, the mere fact that there 
are order $N_c^2$ gluons and only order $N_c$ fermions 
indicates that at leading order in $N_c$ fermions propagate in a 
medium determined by pure gauge theory, with no ``back reaction'' from
the fermions.

Until recently, QCD oriented computer simulations were done 
in the so called quenched (valence) approximation where the fermionic determinant is set 
to unity. Thus, the fermionic ``back reaction'' is eliminated by
hand. Results, perhaps surprisingly, came out quite close to experiment. 
One way to understand this is to argue as in the previous paragraph,
and then assert that this specific property of the 
planar limit holds approximately
also at $N_c=3$. The valence approximation is 
not self-consistent for any finite $N_c$
and does not make up an acceptable field theory.
The main problems occur for massless quarks: 
the fermion determinant 
can be zero exactly for massless quarks  
and there are symmetries obeyed by the massless 
fermion propagator but  
violated by the determinant (anomalies). These properties are important
and get lost when the fermion determinant is replaced by unity. 
If in Nature all quarks were very massive
relative to $\Lambda_{\rm QCD}$, the quantitative success of the 
valence approximation would be easy to understand without
appealing to the large $N_c$ expansion. But, the {\sl up} and {\sl down}
quarks are much lighter than $\Lambda_{\rm QCD}$ and dominate the low energy
physics of strong interactions.

Configurations for which the fermion determinant is exactly zero can be
eliminated, thus redefining the quenched approximation to mean that the
fermion determinant is unity for gauge fields of zero topology and zero
for all other gauge fields. This augmented quenched approximation can
be easily implemented on the lattice if one uses exactly chiral lattice
fermions. Furthermore, simulations in the large $N_c$ limit with massless 
quarks in the fundamental representation should be
restricted to zero topology:
The fermion determinant is zero for non-zero topology and then 
it obviously overcomes the pure gauge action exponent, but 
it is much closer to unity than the exponent of the pure gauge action for zero
topology and therefore can in this case 
be replaced by unity at leading order in $\frac{1}{N_c}$. 
For practical simulations the 
lengths of times are such that lattice gauge configurations in 
one topological sector do not evolve 
into another sector when $N_c$ is large enough (depending on the lattice volume,
$N_c$ larger than order 10 eliminates topology changes). 
Thus, even for massive quarks one ends up working always
at a unique total topology. The other problem one has with the quenched
approximation at finite $N_c$, that of anomalies, 
becomes explicitly an effect of subleading order in
the $\frac{1}{N_c}$ expansion. At finite $N_c$ the inconsistency of the
quenched approximation stemming from the absence of the anomaly 
is reflected by new chiral divergences. These divergences are absent when
fermion loops are included.
But these divergent terms have coefficients
that explicitly vanish in the large $N_c$ limit, validating the 
self-consistency of the 
leading term in the large $N_c$ expansion. 
In short, at $N_c=\infty$, fermions are automatically ``quenched'', and 
the field theoretical framework is intact.

In the context of QCD quenching was introduced for the simple reason that 
computers were not powerful enough to include the fermion determinant, while 
excluding this determinant produced a model that one could simulate in practice. One would
guess therefore that relatively modest computer resources (by the standards
of today) would suffice to numerically solve for, say, the meson spectrum
of large $N_c$ gauge theory. However, going to the large $N_c$ limit increases
computational cost, so additional insights into 
simplifications that occur at infinite $N_c$ 
are needed to find ways to offset this. 

One simplification is that one does not have to worry about attaining the
``thermodynamic limit''-- in the usual sense of the word -- when $N_c$ is 
infinite~\cite{cek3,cek4}: 
When $N_c=\infty$ any Wilson loop expectation value 
on a finite $l^4$ torus does not depend on $l$, 
so long as $l>l_c$, where
$l_c$ is a fixed physical length, of order $\frac{3}{4}~{\it fermi}$ in QCD terms.
Another simplification is afforded by a specific trick, first introduced by
Gross and Kitazawa~\cite{gk} and later adapted to the lattice and tested in two 
Euclidean dimensions~\cite{2d}, which eliminates finite $l$ effects also from meson
propagators. The essence of this trick is to ``force-feed'' a differential of momenta
into the two quark lines that make up the meson propagator. 

A putative puzzle now presents itself threatening to invalidate
the exact independence of meson propagators on $l$. Obviously, these
propagators are strongly affected by spontaneous chiral symmetry breaking
for massless quarks at any $N_c$, including $N_c=\infty$. But, symmetries
tend not to break spontaneously in a finite volume, as can easily be checked
explicitly by solving $\lambda ({\vec \Phi}^2)^2$ at infinite $N$, where $N$
is the number of components of the real scalar 
field ${\vec \Phi}$ and the calculation
is done on a finite four torus~\cite{3dphi}. 
The resolution of the puzzle is simple: unlike
in the case just mentioned, spontaneous chiral symmetry breakdown is induced
by the $N_c\to\infty$ limit even on a finite torus, 
of side $l$, at least so long
as $l>l_c$. Moreover, the value of the traditional order parameter,
the expectation value of the fermion bilinear 
condensate, is $l$ independent in any well defined 
renormalization scheme that in itself is volume independent.

This resolution of the puzzle 
is the central hypothesis of the present paper. We shall present 
numerical evidence verifying it. Also, we shall argue that spontaneous
symmetry breaking is very natural at infinite $N_c$, strengthening insights
developed during the last decade on the connection 
of chiral symmetry breaking with random matrix theory.
In short, spontaneous chiral symmetry breaking occurs at infinite $N_c$ as a
result of sufficient disorder, which seems to always be present
so long as we have confinement, at $l>l_c$, but certainly does not require 
confinement.  

\section {Spontaneous chiral symmetry breaking on the lattice.}

Learning how to exactly preserve chiral symmetry on the lattice has
allowed to separate the understanding of the phenomenon 
of spontaneous chiral symmetry breaking (S$\chi$SB) into two questions.
The first asks whether S$\chi$SB occurs at fixed lattice spacing and by
what mechanism. The second asks whether the effect survives the process
of taking the continuum limit, where the lattice spacing goes to zero.
Because ultraviolet physics seems largely irrelevant for the infrared
phenomenon of S$\chi$SB it is widely believed that the first question
addresses the more fundamental dynamical issues. 

We start our study focusing on the first question. At this stage
our simulations are 
carried out at one -- relatively coarse -- lattice spacing $a$, which,
in QCD terms is about $\frac{1}{8}~{\it fermi}$ and at zero total topology.
Addressing the first question is the main objective of this paper. 
Subsequently, we proceed to the second question and check scaling, which
is our secondary objective here. 
The second question is not fully addressed because we do
not determine the needed mass normalization nonperturbatively. 
In addition, the amount of data collected at smaller lattice spacings 
is relatively small. 

Our quarks are massless and we should only consider
zero topology. However, the technique we use to establish S$\chi$SB 
extends to a prediction at non-zero topology, providing a crosscheck. 
We crosscheck our results against a few simulations
at nonzero topology. Also, since we really define the large $N_c$ limit
by taking $N_c$ to infinity before taking the quark mass to zero, the
restriction to zero topology might be questioned. 
We check whether our
zero quark mass results for the chiral condensate 
are compatible with those at finite quark mass. 

In order to get some 
feeling whether the numerical behavior that we see is reasonable we
compare our methods to the situation in two dimensions, the 't Hooft
model. There the exact value of the condensate in the continuum limit
is known, so we can independently judge how far astray the simulations
might lead us. 

The bilinear fermionic action is described by the massless overlap
Dirac operator~\cite{overlap} $D_o$:
\begin{eqnarray}
&D_o = \frac{1+V}{2}\nonumber\\
&V^{-1}=V^\dagger=\gamma_5 V \gamma_5
={\rm sign}(H_w (M))\gamma_5
\end{eqnarray}
$H_w (M)$ is the Wilson Dirac operator at mass $M$, which we shall choose
as $M=-1.5$. $M$ should not be confused with the bare quark mass $m$, to be introduced
below. 
\be
H_w (M)= \gamma_5 \left [ 
M+4 -\sum_\mu \left ( \frac{1-\gamma_\mu}{2} T_\mu +\frac {1+\gamma_\mu}{2} 
T_\mu^\dagger \right ) \right ]
\ee
The $T_\mu$ matrices are the lattice generators of parallel transport and
depend parametrically and analytically on the lattice links $U_\mu(x)$
which are $SU(N_c)$ matrices at site $x$ associated with the link connecting
site $x$ to site $x+\hat\mu$, where $\hat\mu$ is a unit vector in the
positive $\mu$-direction. 

At finite $N_c$ and with $N_f$ flavors we would have to take into account
a factor of $[\det D_o ]^{N_f}$ influencing the distribution of the gauge
fields. As explained in the introduction, at zero topology and infinite
$N_c$ this factor is replaced by unity. The internal fermion-line propagator,
$\frac{2}{1+V}$ is therefore not needed. For fermion lines continuing
external fermion sources we are allowed to use a slightly different
quark propagator~\cite{overprop,EHN} defined by:
\begin{equation}
\frac{1}{A} =\frac{1-V}{1+V}
\end{equation}
$A=-A^\dagger$ and anticommutes with $\gamma_5$. The spectrum of $A$ 
is unbounded, but is determined by the spectrum
of $V$ which is restricted to the unit circle. One should
think of $A$ as dimensionless, and of $|M|$ as providing
the needed dimension. Up to a dimensionful unit, 
$A$ should be thought of as a lattice realization of 
the continuum massless Dirac operator, $D$:
\begin{equation}
2|M| A \leftrightarrow D=\gamma_\mu \partial_\mu + .....
\end{equation}

A positive, dimensionless, quark mass is added by defining~\cite{overprop,EHN}
\begin{equation}
A(\mu) = A+\mu
\end{equation}
The relation between the bare, properly normalized, quark mass $m$ and
the parameter $\mu$ is $m=2|M|\mu$. This can be used to define a massive
overlap Dirac operator, $D_o (\mu)$, appropriate for the internal fermion
lines making up the fermion determinant~\cite{overprop,EHN}. 

The eigenvalues of $V$ can be written as $-e^{-2\imath \theta}$.
Then the eigenvalues of $A$ are $\imath\tan\theta$. The eigenvalues
of the massless overlap operator $D_o$ are given by
$\imath e^{-i\theta} \sin \theta$. For $|\theta| << 1$ the eigenvalues
of $A$ and $D$ are numerically very close. 

At finite $N_c$ the occurrence of S$\chi$SB manifests itself
mathematically as a lack of commutativity of the limits $V\to\infty$ 
(here $V$ is the lattice volume) and $\mu\to 0$. 
\begin{eqnarray}
&\lim_{\mu\to 0} \lim_{V\to\infty} \frac{1}{V} \langle 
Tr A^{-1} (\mu) \rangle_{N_c , V} = \hat \Sigma \ne 0\nonumber\\
&\lim_{\mu\to 0} \langle Tr A^{-1} (\mu) \rangle_{N_c , V}=0
\end{eqnarray}
$\langle ... \rangle $ means gauge averaging, which includes the
factor $[\det D_o (\mu ) ]^{N_f} ]$ and this is an additional source
of dependence on $\mu$. The subscript ``$N_c ,V$'' 
means that the average is performed
on the torus of volume $V$ at fixed and finite  $N_c$.

As $N_c\to\infty$, $\hat \Sigma$ will diverge, 
but $\Sigma\equiv\frac{\hat\Sigma}{N_c}$ will have a finite limit. 
The hypothesis of this paper is that:
\begin{eqnarray}
&\lim_{\mu\to 0} \lim_{N_c\to\infty} \frac{1}{VN_c} 
\langle Tr A^{-1} (\mu) \rangle_{N_c , V} = \Sigma \ne 0\nonumber\\
&\lim_{\mu\to 0} \langle Tr A^{-1} (\mu) \rangle_{N_c , V}=0
\end{eqnarray}
The second line of the above two sets 
of equations 
is a direct consequence of $\gamma_5 A+ A\gamma_5 =0$.
The notation also implies that $\Sigma$ is independent of the volume $V$, 
meaning that $\lim_{\mu\to 0} \lim_{N_c\to\infty} \frac{1}{N_c} 
\langle Tr A^{-1} (\mu) \rangle_{N_c , V}$ has to come out a linear function of 
$V$ with zero intercept. 
A somewhat superficial (but nevertheless quite convincing) argument supporting this
claim starts by imagining an expansion of $\frac{1}{N_c} 
\langle Tr A^{-1} (\mu) \rangle_{N_c , V}$ in traces of Wilson loops, and then 
invokes large $N_c$ reduction for the latter. However, the argument ignores the question whether
the expansion in loops converges - a substantial issue in particular since 
we intend to take the bare quark mass to zero at the end. 

The quantity 
$\hat\Sigma$, as defined above, diverges
in the quenched approximation for finite $N_c$~\cite{quenchdiv,sharpe}. 
However, the quantity
$\Sigma$, as defined above, stays finite
in the $N_c\to\infty$ limit.
We learn that we need to refine our definition of the large $N_c$ limit
for the case of massless quarks. One needs to take the limits in the 
following order: First one takes $N_c\to\infty$ in the 't Hooft way,
at fixed number of flavors, $N_f \ge 1$, and a fixed mass, $\mu\ne 0$, 
and only subsequently one lets $\mu\to 0$. The first step removes
the dependence on $N_f$. There exist other orderings of limits, in
which the dependence on $N_f$ can be preserved. With our order of limits,
the independence on $N_f$ means that one could obtain the same limit
in yet another way: First one takes $N_c\to\infty$ in the 't Hooft way,
for the quenched theory at a fixed mass $\mu>0$, 
and only subsequently one lets $\mu\to 0$. This is the basic definition of 
the large $N_c$ limit we adopt in this paper. The spectra of the massless
and massive Dirac operator are trivially related in the quenched theory.
Therefore, we shall see
later that if we are careful about what we are doing we can get away by taking 
the large $N_c$ limit of the quenched theory at $\mu=0$ directly; 
what exactly we do and why will
become clear only after we speak in greater 
detail about infrared divergences coming from quenching.

\section {Random Matrix Theory (RMT) and S$\chi$SB}

Clearly, the noncommutativity of limits described in
the previous section implies nontrivial crossover
effects in the dependence of the quantity 
$\langle Tr A^{-1} (\mu) \rangle_{N_c , V}$
on $\mu, V$ and $N_c$. This rapidly becomes a numerical burden, since
isolating the right regime might require
extremely large values of $V$ (at fixed $N_c$) or $N_c$ (at fixed $V$).
Fortunately, due to the successful efforts of many workers~\cite{vw}, 
the physics of the crossover regime is quite well 
understood. As is usually the case, the
crossover is largely universal, and, in this
case, there exists a representative of its
universality class given by a Gaussian random matrix model, introduced 
by Shuryak and Verbaarschot~\cite{sv}.

For $N_f \ge 1$ and finite $N_c$ the crossover regime controlling
the pair of limits $\mu\to 0$ and $V\to \infty$ can be described by
traditional methods, employing chiral Lagrangians~\cite{chlan}. At zero topology,
the dependence on $\mu$ and $V$ in the limit $\mu\to 0$ with $\mu V$
held fixed is expressed by a matrix integral over a $U(N_f )$ matrix
$U_0$. The role of the QCD partition function is played by $Z_{\rm eff}$:
\begin{equation}
Z_{\rm eff} (z) =\int dU_0 e^{z\Re Tr U_0},~~z=\mu V {\hat \Sigma_{\rm eff}}
\end{equation}
The constant $\hat \Sigma_{\rm eff}$ is a parameter inherited 
from the complete theory. This is often quoted in continuum contexts, 
but also holds on the lattice. It is in the latter sense that we write
the formula here. Under the conditions stated above,  
$\hat \Sigma_{\rm eff}= \hat\Sigma $. 

The quenched case is formally obtained by letting $N_f$ approach zero.
This additional limit is not very well defined since theories with
a fractional $N_f$ are non-local and therefore may develop additional
singularities and needs special care~\cite{damsplit}. The quenched case can also be 
studied by the ``supersymmetric 
method'', which appears more reliable~\cite{quenchdiv,quenchcond}.
The upshot of such an analysis
is that there are infrared renormalization effects from fluctuating
fields that cannot be neglected. These effects can be absorbed into
a redefinition of $\hat \Sigma_{\rm eff}$, which now becomes dependent
on $V$, complicating the formulation of the crossover effect~\cite{quenchcond}.
Actually,
$\hat\Sigma (V)$ diverges as $V\to\infty$, but the divergence is probably
weak enough to allow a separation of scales which maintains the 
applicability of the matrix integral formula for $Z_{\rm eff}$. 
It is somewhat unclear how solid this last claim is for finite $N_c$.
However, things simplify again if one also expands in $\frac{1}{N_c}$.
The volume dependence of  $\frac{1}{N_c}\hat\Sigma (V)$ enters then
order by order in the $\frac{1}{N_c}$ expansion. The leading term
is volume independent, $lim_{N_c\to\infty} \frac{1}{N_c}\hat\Sigma (V)=\Sigma$
and the first subleading term has a logarithmic dependence on $V$.
Thus, the dependence on $V$ disappears also in the quenched case if we
take the large $N_c$ limit. In short, the large $N_c$ limit -- 
as we defined it -- 
of ordinary quenched theories is free of infrared divergences and coincides
with the large $N_c$ limit of theories with finite and fixed $N_f\ge 1$.

The formula for $Z_{\rm eff}$ would hold in any fundamental model 
that has the same flavor symmetries and breaking pattern as QCD.
However, the case of QCD enjoys one special property that we have
not yet exploited, namely that fermions enter the action only bilinearly.
This property has been a past source of other important results, like
the geometrical structure of anomalies and mass inequalities.
In our context, the bilinear structure of the fermionic action, when combined
with chiral effective Lagrangian techniques leads to chiral 
random matrix theory. This produces a more QCD specific and a more detailed
understanding of the crossover regime. 

The main result of chiral random matrix theory concerns the 
eigenvalues of the operator $A$, $\imath\lambda$~\cite{vw,dn}. 
Because of the symmetry
of the spectrum, and since we are working at zero topology,
we can focus on $\lambda >0$, and order the eigenvalues
as $\lambda_1 \le \lambda_2 \le \lambda_3 \le .... \le \lambda_K$, where
$K=2N_c L^4$. As the quantity $N_c L^2$ increases and goes through a certain
threshold more and more of the low eigenvalues attain universal distributions
in terms of random variables $z_k$, with $z_k=\lambda_k\Sigma N_c L^4$
\footnote{There have been confusing statements in the literature regarding the question
whether the chiral effective Lagrangian (say, at zero quark mass) based approach
to finite volume effects with toroidal compactification indeed implies the
specific RMT forms for the distributions of individual smallest eigenvalues. 
We accept the reasoning of the recent paper~\cite{akedam}, showing 
that the answer to this question is yes.}.
The $z_k$'s are the square roots of the ordered
eigenvalues of a matrix $C^\dagger C$, 
where $C$ is a complex random matrix of size $R$, distributed as a Gaussian:
\begin{equation}
d\mu (C) ={\cal N} e^{-RTr C^\dagger C}\prod_{1\le i\le j \le R} d^2 C_{ij}
\end{equation}
At fixed lattice spacing the product $N_c L^2$ can be increased by increasing
$N_c$ at fixed $L$. This produces the same effect as increasing $L$ at fixed
$N_c$. Once $\lambda_1$ and $\lambda_2$ are 
determined to be jointly distributed 
in accordance with random matrix theory, the scaling parameter 
$\Sigma N_c L^4$ can be numerically extracted. This shows in detail how 
large $N_c$ reduction works in this case, since the parameter $\Sigma$ is 
defined {\sl a priori} in the limit in which $N_c$ and $L$ are simultaneously
infinite.

Thus, our numerical task is as follows: 
We first need to show that it is enough
to increase $N_c$ at fixed $L$ to cause $\lambda_{1,2}$ to become distributed
according to random matrix theory. Once this is established 
we should extract $\Sigma$ for different
values of the fixed parameter $L$, and show that the different
numbers are the same, and therefore $L$-independent. 
Subsequently, we should check whether more traditional ways to
extract $\Sigma$ lead to consistent results.
We have carried out the required numerical work and our results support
the hypothesis of this paper. Details will be presented in 
section~\ref{overview}.

Encouraged by these results, we now provide a somewhat different viewpoint.
Until now we relied mainly on theoretical 
results that were obtained at finite $N_c$. 
At finite $N_c$ it is unlikely that the entire Dirac operator is a random
matrix in any precise sense. It is only a tiny 
fraction of its lowest eigenvalues
that are distributed as if they were the eigenvalues of a random Dirac matrix.
One could imagine framing the analysis as follows: Assume that one does the
integral over all gauge fields subjected to constraints that fix
$Tr A^{-2k}$ for all $K\ge k\ge 1$. This seems to define a distribution
of $\{\lambda_1 ,\lambda_2,.....,\lambda_K\}$. For finite $N_c$ the number $K$
is a finite fraction of the total number of gluonic degrees of freedom.
The totality of $\{\lambda_k; K\ge k \ge 1\}$ contains as much information
as a full fledged (highly nonlocal)
quantum field and therefore its fluctuations are not expected to 
have a structure largely independent of the detailed dynamics.   
It is therefore unlikely that the distribution of  
$\lambda_1 ,\lambda_2,.....,\lambda_K$ have eigenvalue repulsion of the
universal type for all adjacent eigenvalues. 

However, at infinite $N_c$,
$K$ is a vanishing fraction of the total number of degrees of freedom
since $K$ is linear in $N_c$ and there are order $N_c^2$ 
gauge degrees of freedom.
It is now more likely that the simplest possibility holds: as the size of $A$
goes to infinity the repulsion between all adjacent 
eigenvalues is of the universal
type. At zero topology $A$ has the structure
\begin{equation}
A=\pmatrix { 0 & C\cr -C^\dagger & 0 }
\end{equation}
$C$ is a square matrix of size $K\times K$
and depends parametrically on the fluctuating gauge fields.
(We assume that $\gamma_5=\pmatrix{1&0\cr 0&-1}$).
As such, the distribution of $C$ inherits an exact 
invariance under gauge transformations. The volume is finite, but this
is not reflected by $C$ being sparse in any precise sense. It is true
that at infinite $L$ entires of $C$ that are far in site distance
would be small: but this is not very relevant for finite $L$. Thus,
the $C$-block of $A$ is probably well described by a random matrix
model with $U(K)\times U(K)$ invariance. By ``well described'' we mean
that after ``unfolding'' (rescaling all eigenvalues by a local average level
density) the spectrum of $A$ will be universal in the limit $N_c\to\infty$.
While the local average level density is gauge theory specific, 
bears all the dynamics, and in many respects is constrained by ordinary field
theory universality, the local microscopic level distribution is 
unaffected by the gauge theory dynamics and obeys random matrix universality. 
We believe that many more matrix observables in large $N_c$ gauge
theory exist, obeying such dual universalities. The Dirac
operator is, we conjecture, a nice example of this concept. 
These conjectures lead to an effective weakening of the
importance of space-time locality at infinite $N_c$.

It has been observed long ago that 
S$\chi$SB is equivalent to the statement
that the density of eigenvalues of 
$A$ is nonzero at very small eigenvalues,
contrary to the free case, where it 
vanishes as $|\lambda|^3$ (in four dimensions)~\cite{bc}.
At finite four-volume and finite $N_c$ this amounts to
a roughly equally spaced spectrum near zero with eigenvalues and spacings
of order $\frac{1}{N_c L^4}$. This kind of spectrum is obtained generically 
in RMT where eigenvalues repel and the spectrum tends to be rigid, 
with small fluctuations in the total number of 
eigenvalues in a given interval. 
Thus, RMT naturally provides  S$\chi$SB. All one needs is 
that the randomness be
sufficiently strong, and that the asymptotic order of magnitude
of eigenvalues and spacings for large $R$, where $R$ 
is the dimension of the the vector space the random 
matrix acts on, be given by $\frac{{\rm Const.}}{R}$.
This asymptotic behavior is very natural, since the basic
unitary matrix $V$ is unitary, 
with a spectrum restricted to the unit circle.  
This is what seems to happen in the $l > l_c$ phase of planar
QCD on a torus of side $l$, where all large Wilson loops are expected to 
obey an area law. In short, at infinite $N_c$, the occurrence of 
S$\chi$SB already at 
finite four volume is a plausible consequence of the randomness of the
gauge fields.  

The relevant formula from chiral RMT for our numerical work are the
distributions of the two lowest eigenvalues and the distribution of
their ratio in the $Q=0$ and $Q=1$ topological sector~\cite{dn}. The distribution
of the lowest scaled eigenvalue $z_1$ 
and the second scaled eigenvalue $z_2$ in the $Q=0$ topological sector
are given by
\be
p_1(z_1) = \frac{1}{2} z_1 e^{ - \frac{z_1^2}{4}}
\ee
\be
p_2(z_2) = \frac{1}{4} e^{-\frac{z_2^2}{4}} z_2 \int_0^{z_2} du u 
\bigl [ I_2^2(u) - I_1(u) I_3(u) \bigr ]
\ee
The distribution of the ratio $r=\frac{z_1}{z_2}$ in the $Q=0$
topological sector is given by
\be
p(r) = \frac{1}{4} \frac {r}{(1-r^2)^2} \int_0^\infty du
e^{-\frac{u^2}{4(1-r^2)}} u^3 \bigl [ I_2^2(u) - I_1(u) I_3(u) \bigr]
\ee
The distribution
of the lowest scaled eigenvalue $z_1$ 
and the second scaled eigenvalue $z_2$ in the $Q=1$ topological sector
are given by
\be
p_1(z_1) = \frac{1}{2} e^{-\frac{z_1^2}{4}} z_1 I_2(z_1)
\ee
\begin{eqnarray}
p_2(z_2) = \frac{1}{4} e^{-\frac{z_2^2}{4}} &&
\int_0^{z_2} \frac{dz_1}{z_1}  \Bigl \{
z_2 I_2(z_2) (z_2^2-z_1^2)\bigl [ 
I_1^2(\sqrt{z_2^2-z_1^2}) - I_0(\sqrt{z_2^2-z_1^2})I_2(\sqrt{z_2^2-z_1^2})
\bigr ] \cr
&+& z_2^2 I_1(z_2) \sqrt{z_2^2-z_1^2}\bigl [ 
I_0(\sqrt{z_2^2-z_1^2})I_3(\sqrt{z_2^2-z_1^2}) 
- I_1(\sqrt{z_2^2-z_1^2})I_2(\sqrt{z_2^2-z_1^2}) \bigr] \cr
&+& z_2^3 I_0(z_2) \bigl [ 
I_2^2(\sqrt{z_2^2-z_1^2}) - I_1(\sqrt{z_2^2-z_1^2})I_3(\sqrt{z_2^2-z_1^2})
\bigr]
\Bigr \}
\end{eqnarray}
The distribution of the ratio $r=\frac{z_1}{z_2}$ in the $Q=1$
topological sector is given by
\begin{eqnarray}
p(r) = \frac{1}{4r} && \int_0^\infty dz z^3 e^{-\frac{z^2}{4}}
\Bigr \{ (1-r^2) I_2(z) \bigl [ 
I_1^2(z\sqrt{1-r^2}) - I_0(z\sqrt{1-r^2})I_2(z\sqrt{1-r^2}) \bigr] \cr
&+& \sqrt{1-r^2} I_1(z) \bigl [ 
I_0(z\sqrt{1-r^2})I_3(z\sqrt{1-r^2}) - I_1(z\sqrt{1-r^2})I_2(z\sqrt{1-r^2})
\bigr] \cr
&+& I_0(z) \bigl [ 
I_2^2(z\sqrt{1-r^2}) - I_1(z\sqrt{1-r^2})I_3(z\sqrt{1-r^2}) \bigr]
\Bigr \}
\end{eqnarray}
The integrals involving modified Bessel functions
were numerically evaluated to a high precision and accurate plots 
of the various distributions were generated.

\section {Essentials of the lattice formulation.}

We used the simplest, single plaquette, pure gauge Wilson action, given by
\begin{eqnarray}
S=\frac{\beta}{4N_c}\sum_{x,\mu\ne\nu} Tr[ U_{\mu,\nu}(x)
+U_{\mu,\nu}^\dagger (x) ] \\
U_{\mu,\nu}(x)=U_\mu (x) U_\nu (x+\mu) U_\mu^\dagger (x+\nu) 
U_\nu^\dagger (x)
\end{eqnarray}
We define $b=\frac{\beta}{2N_c^2}=\frac{1}{g^2N_c}$ and take 
the large $N_c$ limit with $b$ held fixed. As usual, $b$ determines
the lattice spacing $a$. 
The lattice is a symmetric torus
of side $L$ in lattice units. 
The gauge fields are periodic. $x$ is a four component integer
vector labeling the site, and $\mu$ either labels a direction or denotes
a unit vector in the $\mu$ direction. 
The link matrices $U_\mu(x)$ 
are in $SU(N_c)$. The total topological charge, using the ``overlap'' 
definition~\cite{overtop}
with mass parameter $M$ set to $-1.5$ is denoted by the integer $Q$. 
\begin{equation}
Q=\frac{1}{2} Tr \left [ H_w (M) \right ]
\end{equation}
We generated $SU(N_c)$ gauge field configurations at several values of $b$.
The evaluations of fermionic observables in a 
given gauge field background were separated by $25$ updates.
In our terminology, one update of the lattice corresponds to
one Cabibbo-Marinari heatbath update for each one of the 
$\frac{N_c(N_c-1)}{2}$ $SU(2)$ subgroups of $SU(N_c)$ for each link, 
followed by one full $SU(N_c)$
overrelaxation pass over the entire lattice~\cite{cek4}. 
Thermalization at a fixed $L$, $N_c$ and $Q$ was achieved by performing
a total of $500$ updates. The starting configuration 
at $Q=0$ was 
\begin{eqnarray}
U_1(i_1,i_2,i_3,i_4) &=& \cases { 1 & for $1 \le i_1 < L$,
$1 \le i_2,i_3,i_4 \le L$ \cr
z_1 & $i_1 = L$, $1 \le i_2,i_3,i_4 \le L$ } \cr
U_2(i_1,i_2,i_3,i_4) &=& \cases { 1 & for $1 \le i_2 < L$,
$1 \le i_1,i_3,i_4 \le L$ \cr
z_2 & $i_2 = L$, $1 \le i_1,i_3,i_4 \le L$ } \cr
U_3(i_1,i_2,i_3,i_4) &=& \cases { 1 & for $1 \le i_3 < L$,
$1 \le i_1,i_2,i_4 \le L$ \cr
z_3 & $i_3 = L$, $1 \le i_1,i_2,i_4 \le L$ } \cr
U_4(i_1,i_2,i_3,i_4) &=& \cases { 1 & for $1 \le i_4 < L$,
$1 \le i_1,i_2,i_3 \le L$ \cr
z_4 & $i_4 = L$, $1 \le i_1,i_2,i_3 \le L$ } 
\end{eqnarray}
where $z_\mu$, $\mu=1,2,3,4$ are randomly chosen members of $Z_{N_c}$.

For a $Q=1$ configuration we started with all links defined by the ``uniform'' instanton
configuration below:
\begin{eqnarray}
U_1^{i,j}(i_1,i_2,i_3,i_4) &=& \cases { 
e^{-i\frac{2\pi(i_2-1)}{L}}; & for $i=j=1$; $1 \le i_2,i_3,i_4 \le L$; $i_1=L$ \cr
e^{i\frac{2\pi(i_2-1)}{L}}; & for $i=j=3$; $1 \le i_2,i_3,i_4 \le L$; $i_1=L$ \cr
1; & for $i=j \ne (1,3)$; $1 \le i_1,i_2,i_3,i_4 \le L$ \cr
0; & elsewhere } \cr
U_2^{i,j}(i_1,i_2,i_3,i_4) &=& \cases { 
e^{i\frac{2\pi(i_1-1)}{L^2}}; & for $i=j=1$; $1 \le i_1,i_2,i_3,i_4 \le L$\cr
e^{-i\frac{2\pi(i_1-1)}{L^2}}; & for $i=j=3$; $1 \le i_1,i_2,i_3,i_4 \le L$\cr
1; & for $i=j \ne (1,3)$; $1 \le i_1,i_2,i_3,i_4 \le L$\cr
0; & elsewhere } \cr
U_3^{i,j}(i_1,i_2,i_3,i_4) &=& \cases { 
e^{-i\frac{2\pi(i_4-1)}{L}}; & for $i=j=2$; $1 \le i_1,i_2,i_4 \le L$; $i_3=L$ \cr
e^{i\frac{2\pi(i_4-1)}{L}}; & for $i=j=3$; $1 \le i_1,i_2,i_4 \le L$: $i_3=L$\cr
1; & for $i=j \ne (2,3)$; $1 \le i_1,i_2,i_3,i_4 \le L$\cr
0; & elsewhere } \cr
U_4^{i,j}(i_1,i_2,i_3,i_4) &=& \cases { 
e^{i\frac{2\pi(i_3-1)}{L^2}}; & for $i=j=2$; $1 \le i_1,i_2,i_3,i_4 \le L$\cr
e^{-i\frac{2\pi(i_3-1)}{L^2}}; & for $i=j=3$; $1 \le i_1,i_2,i_3,i_4 \le L$\cr
1; & for $i=j \ne (2,3)$; $1 \le i_1,i_2,i_3,i_4 \le L$\cr
0; & elsewhere } 
\end{eqnarray}
No tunneling during our thermalization cycle ever occurred, so the
equilibrium configurations also have $Q=1$. As is well known, during a
simulation, the ``overlap'' topological charge will change only
extremely rarely (far beyond what is observable in practice) whenever all
gauge configurations produce a hermitian Wilson matrix $H_w$ that has
a finite gap around zero. At large $N_c$ the gauge configurations come
naturally out preserving this gap~\cite{2d} and this explains the
absence of tunneling events. In turn, the gap in the spectrum of $H_w$
is a consequence of a gap in the eigenvalue distribution of the
parallel transporters round single plaquettes. The latter have a gap
because we are working in a phase which is disconnected from the
regime of lattice strong coupling by a phase transition where this gap
forms. The existence of a gap in $H_w$ is of great practical value
when dealing with the overlap Dirac operator because it eliminates the
need to deal separately with the subspace of $H_w$ associated with
very small eigenvalues in absolute magnitude. Technically, we do not
need to ``project out'' these states when we use the pole approximation
to evaluate the action of the unitary matrix $V$~\cite{single}.  Also, a robust gap
in $H_w$ makes the definition of $Q$ practically unambiguous (in the
sense that one can vary the parameter $M$ in a reasonable range and
the assignment of topological charge will not change).  Note that at
large $N_c$ the computer cost balance between the single pass~\cite{single} and
double pass~\cite{double} switches relative to $N_c=3$~\cite{twchiu} 
and, typically, the single
pass version, with Zolotarev coefficients~\cite{zol}, is more efficient.  Single
pass with too large $N_c$ would be prohibitive because of memory
considerations, but, in practice, we did not need to deal with this
issue.

\section{Overview of numerical work}\label{overview}

\begin{table}
\caption{\label{tab1} List showing the number of configurations
used at different $L$, $N_c$ and $Q$ for the analysis. The table
also displays results
for the average of the ratio of the first eigenvalues to the
second eigenvalue divided by its RMT value 
and the estimates of the condensate from the
first and second eigenvalues.} 
\begin{ruledtabular}
\begin{tabular}{rrccrccc}
$b$ & $L$ & $N_c$ & $Q$ & No. of conf. & 
$\frac{\langle {\lambda_1}/{\lambda_2}
\rangle}{ \langle {\lambda_1}/{\lambda_2}
\rangle_{\rm RMT}}$ 
& $\Sigma_1^{1/3}$ & $\Sigma_2^{1/3}$ \\
\hline
0.346 & 9&11&1& 96&0.99(3)&0.1551(20)&0.1553(10)\\
0.350 & 6&13&0&436&1.99(2)&0.1065(05)&0.1356(04)\\
0.350 & 6&17&0&511&1.65(2)&0.1108(08)&0.1333(05)\\
0.350 & 6&23&0&446&1.11(2)&0.1421(12)&0.1485(06)\\
0.350 & 6&29&0&599&1.06(2)&0.1419(10)&0.1450(05)\\
0.350 & 6&37&0&287&1.01(3)&0.1458(15)&0.1464(07)\\
0.350 & 6&37&1&192&1.05(2)&0.1440(13)&0.1469(07)\\
0.350 & 6&43&0&292&1.02(3)&0.1401(14)&0.1406(07)\\
0.350 & 7&17&0&346&1.07(3)&0.1376(13)&0.1412(06)\\
0.350 & 7&19&0&315&0.99(3)&0.1407(14)&0.1408(06)\\
0.350 & 7&23&0&440&1.01(3)&0.1415(12)&0.1426(05)\\
0.350 & 7&29&0&348&1.02(3)&0.1441(13)&0.1454(06)\\
0.350 & 7&29&1&288&1.04(2)&0.1403(09)&0.1426(05)\\
0.350 & 8&13&0&310&1.04(3)&0.1352(13)&0.1371(06)\\
0.350 & 8&17&0&270&1.04(3)&0.1384(15)&0.1399(07)\\
0.350 & 8&23&0&257&1.03(3)&0.1398(14)&0.1413(07)\\
0.350 & 8&23&1&288&1.01(2)&0.1418(10)&0.1426(06)\\
0.350 &10&11&0& 64&1.05(6)&0.1313(27)&0.1333(17)\\
0.355 & 8&23&0&288&1.03(3)&0.1192(13)&0.1205(05)\\
0.355 & 9&17&0&288&1.04(3)&0.1204(12)&0.1220(06)\\
0.355 &10&13&0&288&1.09(3)&0.1118(11)&0.1160(05)\\
0.3585& 9&17&0&336&1.10(3)&0.1083(10)&0.1121(04)\\
0.3585& 9&23&0&300&1.03(3)&0.1062(11)&0.1078(05)\\
\hline
\end{tabular}
\end{ruledtabular}
\end{table}

We started our study at a lattice gauge coupling 
$b=0.350$ where the bare 't Hooft coupling is given by
$g^2 N_c=\frac{1}{b}$. At this lattice coupling
a lattice size of $L=6$ is very close to critical~\cite{cek4}. Thus,
$a$, the lattice spacing, 
is approximately given by $a=\frac{l_c}{6}$.
To get some physical idea of this size in QCD terms consider
$N_c=3$ and assume $l_c=\frac{1}{T_c}$, where $T_c$ is the
pure gauge finite temperature gauge transition. This means that
in QCD terms 
$a=\frac{1}{6T_c}\approx\frac{0.75~{\it fermi}}{6}=\frac{1}{8}~{\it fermi}$. 
We kept $b$ fixed at this value and simulated the pure gauge sector
at $L=6,7,8$. From the continuum viewpoint this provides for increasing
physical volumes at fixed physical lattice spacing.  

At each $L$ we simulated several increasing 
values of $N_c$, until we reached a regime where the calculated
$\lambda_1\le\lambda_2$ had a joint RMT distribution. This
question was addressed in a scale independent way by
considering the distribution of the ratio $r=\frac{\lambda_1}{\lambda_2}$,
$p(r)dr$. The random variable $r$ takes values in the segment $(0,1)$
and the behavior of the distribution near $1$ reflects the typical RMT
eigenvalue repulsion. The behavior of the distribution near $0$ is
dominated by the universal features of ``edge'' behavior in RMT. 

The point of looking at $p(r)$ is that the defining property of randomness
in the context of random matrix theory is eigenvalue repulsion. The tail
of the distribution of $p(r)$ near $r=1$ reflects this repulsion best.
As $N_c$ increases at fixed $L$, the lowest eigenvalues of $-A^2$ enter the regime
where they ought to be described by chiral random matrix theory if there
is S$\chi$SB. They enter the regime by being gradually squeezed closer together
by the increase in $N_c$ (recall that the matrix $V$ is unitary, 
so its eigenvalues are restricted to the unit circle), 
until their distribution is totally governed by
eigenvalue repulsion. This means that we expect that the approach to RMT will be by
distributions $p(r)$ which show a gradual depletion close to $r=1$, approaching the
RMT prediction from above in that region, and therefore, to compensate for the
fixed normalization, from below in the region of small $r$ close to zero. This means
that the average of $r$ will approach its RMT limit from above as $N_c$ increases.

Therefore one expects that once the empirical average of $r$ is close to its RMT limit, the
entire distribution $p(r)$ will fall into statistical agreement with the RMT prediction. 
This expectation was confirmed by our data. 
Note that using the random variable $\frac{1}{r}$ instead of $r$ is ill advised, since 
then the behavior at $r=0$ gets emphasized, and the variance of $\frac{1}{r}$ diverges. 
We found that we needed at least several hundred gauge field configurations
to be able to make a reasonable determination that the RMT regime 
has been entered. In general, the fluctuations are quite large
and our statistics are relatively low. To be safe, we feel one should multiply
our errors by a factor of 2 to 3 in order to get some reasonably reliable
confidence intervals. To check whether we are not missing some hidden effect here,
we went to non-zero topology and checked whether the method of random matrix
theory works similarly and consistently there. The results at nonzero topology
indeed behaved as expected.

To this point we have not introduced a mass yet. To see how the large
$N_c$ limit sets in numerically we looked for a quantity that approaches
$\Sigma$ as $\mu\to 0$. The simplest such quantity is the 
bilinear condensate $\langle\bar\psi\psi\rangle(\mu)$,
but, for nonzero $\mu$ this quantity diverges quadratically when $a$
goes to zero. This effect indicates that we may need to subtract
some large numbers from the measured value, enhancing the noise/signal
ratio. We preferred a quantity with a milder behavior in the ultraviolet.
We chose the zero momentum pion-pion scattering matrix element. 
At nonzero but small quark mass and at leading order 
in the $\frac{1}{N_c}$ expansion it is given by
the gauge average of $Tr \left ( \frac{1}{-A^2+\mu^2} \right )^2$. One has then:
\begin{equation}
\lim_{\mu\to 0} \lim_{N_c\to\infty} \frac{2\mu^3}{N_c V} 
Tr \left ( \frac{1}{-A^2+\mu^2} \right )^2 =\Sigma
\end{equation}
While the quantity we are looking at would still have 
a divergent contribution as $a$ goes to zero, 
that divergence is only logarithmic, and not threatening numerically. 
Moreover, the numerical effort involved in evaluating the
latter quantity is equal to that involved in computing the former. 
We found consistent results, but also learned that the 
random matrix method is by far more
accurate as a numerical tool to establish both the presence of S$\chi$SB and
the value of the condensate. 

The comparison between the direct determination of the condensate and the
one via random matrix theory left us somewhat uneasy because the direct determination,
in itself, did not provide convincing evidence for S$\chi$SB. To be sure, 
if one first postulated S$\chi$SB,
the numbers were consistent with the findings using random matrix theory.
Nevertheless, to reassure ourselves that nothing was wrong, 
we looked at identical issues in the 't Hooft model. (Although the model is
in two dimensions, infinite $N_c$ brings it closer to the planar four dimensional
case, because spontaneous symmetry breaking of continuous 
symmetries is now an open option.) In the 't Hooft model we can 
calculate analytically the value of the
condensate in the continuum limit. This provides an independent check showing
that our methods do yield correct results and indeed one can trust the random
matrix method even when direct measurements might appear inconclusive by themselves.

We continued the zero mass analysis to finer lattices, looking at
couplings $b=0.355,0.3585$. These couplings are chosen to be almost equal 
to the critical bare coupling $b_c(L)$ for sizes $L=7,8$~\cite{cek4}. Thus, one expects the
lattice quantity $\Sigma^{\frac{1}{3}} (b_c (L))$ to 
go as $\frac{1}{L}$ so long as the variation
of the normalization constant for the pseudoscalar density varies slowly 
over the range. 

We can estimate from 1-loop continuum perturbation theory how much one expects
the renormalization constant to contribute to the variation of $\Sigma^{\frac{1}{3}}$ 
with lattice spacing. To get a finite continuum condensate in some renormalization 
scheme one needs to multiply the lattice condensate by an appropriate constant $Z_S (\lambda)$.
Slightly modifying known results, we have to 1-loop:
\be
Z_S (\lambda) = 1+ \lambda[\frac{3}{(4\pi)^2} \log (a\mu) + c]
\ee
Here, $\mu$ is a renormalization point and $a$ is the lattice spacing. 
$c$ is a scheme dependent constant~\cite{harris}, irrelevant for the estimates below.
Thus,
\be
\frac{\partial Z_S }{\partial a}\Bigg|_\lambda = \frac{3}{(4\pi)^2b} \left ( \frac{1}{a}\right )
\ee
For us the lattice spacing goes as $\frac{1}{L}$, therefore
$\frac{\delta a}{a}\sim\frac{1}{7}$,
and $b\sim 0.355$ results in a $\delta Z_S\sim 0.008$. 
The contribution coming from $\delta\lambda$ also is small.
The variation $\delta Z_s$ is away
from a value of order unity, so contributes to the change in
the condensate at the level of one percent.
But, the condensate's non-anomalous scale dependence is $a^3$ which would amount
to a non-anomalous change of about 60 percent. Within our accuracy the anomalous contribution
coming from $Z_S$ is therefore too small to be detectable. Thus, scaling will be verified
just by observing the lattice $\Sigma^{\frac{1}{3}}$ go approximately as $\frac{1}{L}$.

\section{Numerical results in four dimensions}

\subsection{Analysis based on chiral random matrix theory}

The two lowest non-zero eigenvalues of interest of the massless Dirac
operator were computed using the Ritz functional method~\cite{ritz} on
$D_oD_o^\dagger$. We worked in a fixed chiral sector and computed
the two lowest eigenvalues of $D_oD_o^\dagger$.  When $Q=1$ we
made sure to work in the chiral sector where there is no zero mode.

The two lowest eigenvalues were used to obtain an estimate of the
chiral condensate in the large $N_c$ limit. Let $\lambda_1\le\lambda_2$ be the
two lowest eigenvalues in a fixed background gauge configuration. For large enough
$N_c$ the 
probability distribution of the ratio $r=\frac{\lambda_1}{\lambda_2}$ is given by
a universal function that only depends on $Q$ and is defined by chiral random matrix theory.
If the probability distribution of the ratio agrees with the prediction from
the chiral random matrix theory, one is confident that the distributions of
the scaled eigenvalues $z_i=\lambda_i \Sigma_i N_c L^4$
should also agree with the predictions from chiral random matrix theory. If this 
confidence is well based we must find that $\Sigma_1$ equals $\Sigma_2$. 

For a given $L$,~$N_c$ and $Q$ we defined $\Sigma_i$ as 
the numbers that made the
observed average of $z_i$ agree with the numbers from chiral random matrix theory with a Gaussian
weight as explained before.
We checked to see if $\Sigma_1=\Sigma_2$ and if the detailed scaled distribution agreed
with chiral random matrix theory predictions. 
The list of different values of $L$, $N_c$, $b$ and $Q$ where we 
did simulations is shown in table~\ref{tab1}. For most cases, we 
see general agreement with
chiral random matrix theory.

\begin{figure}
\epsfxsize = \textwidth
\centerline{\epsfbox{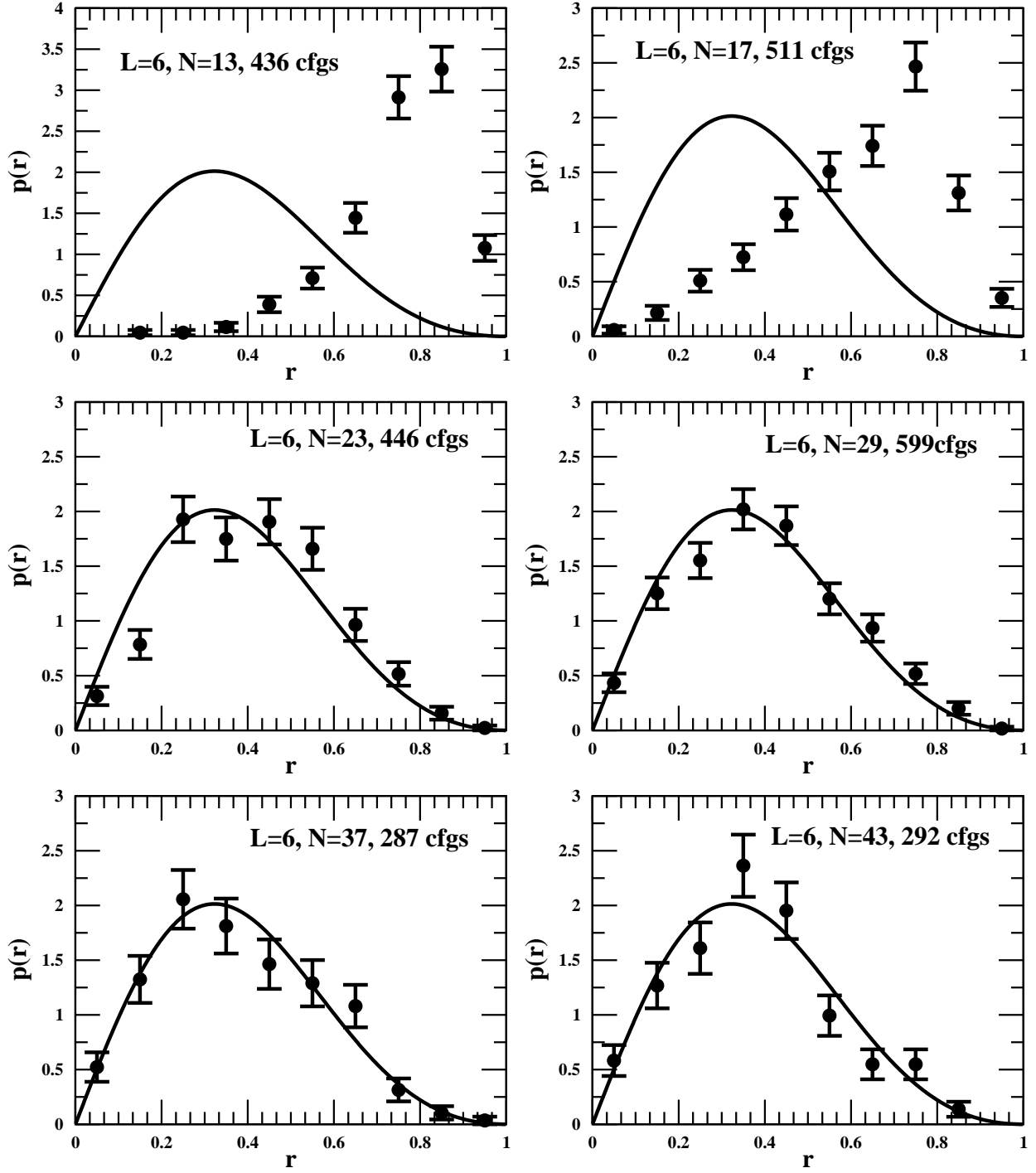}}
\caption{The distribution of the ratio of the first eigenvalue to
the second eigenvalue in the zero topological sector for various $N_c$ 
at $L=6$ and $b=0.350$ is compared with the prediction from chiral RMT in
four dimensions.
}
\label{ratiol6q0b350}
\end{figure}

Fig.~\ref{ratiol6q0b350} compares our data to
chiral random matrix theory for the distribution $p(r)$ in the
$Q=0$ sector for lattice size $6^4$ and coupling $b=0.350$ as $N_c$ increases. 
One should keep in mind that what we are plotting
are histograms, a collection of integers obtained by binning the
data. The integer associated with a given bin undergoes Poissonian
fluctuations, so the standard deviation is equal to the square root
of the same integer. For this reason, the more populated a bin is the larger
is the absolute error bar associated with it. The distributions in the
different bins are somewhat correlated, but there are sizable fluctuations from bin
to bin and this is expected. Even within our limited statistics it is
obvious that a dramatic change in the distribution of the ratio occurs
between $N_c=17$ and $N_c=23$, and that by $N_c=29$ the distribution
has stabilized, essentially reproducing the infinite $N_c$ limit. 

\begin{figure}
\epsfxsize = \textwidth
\centerline{\epsfbox{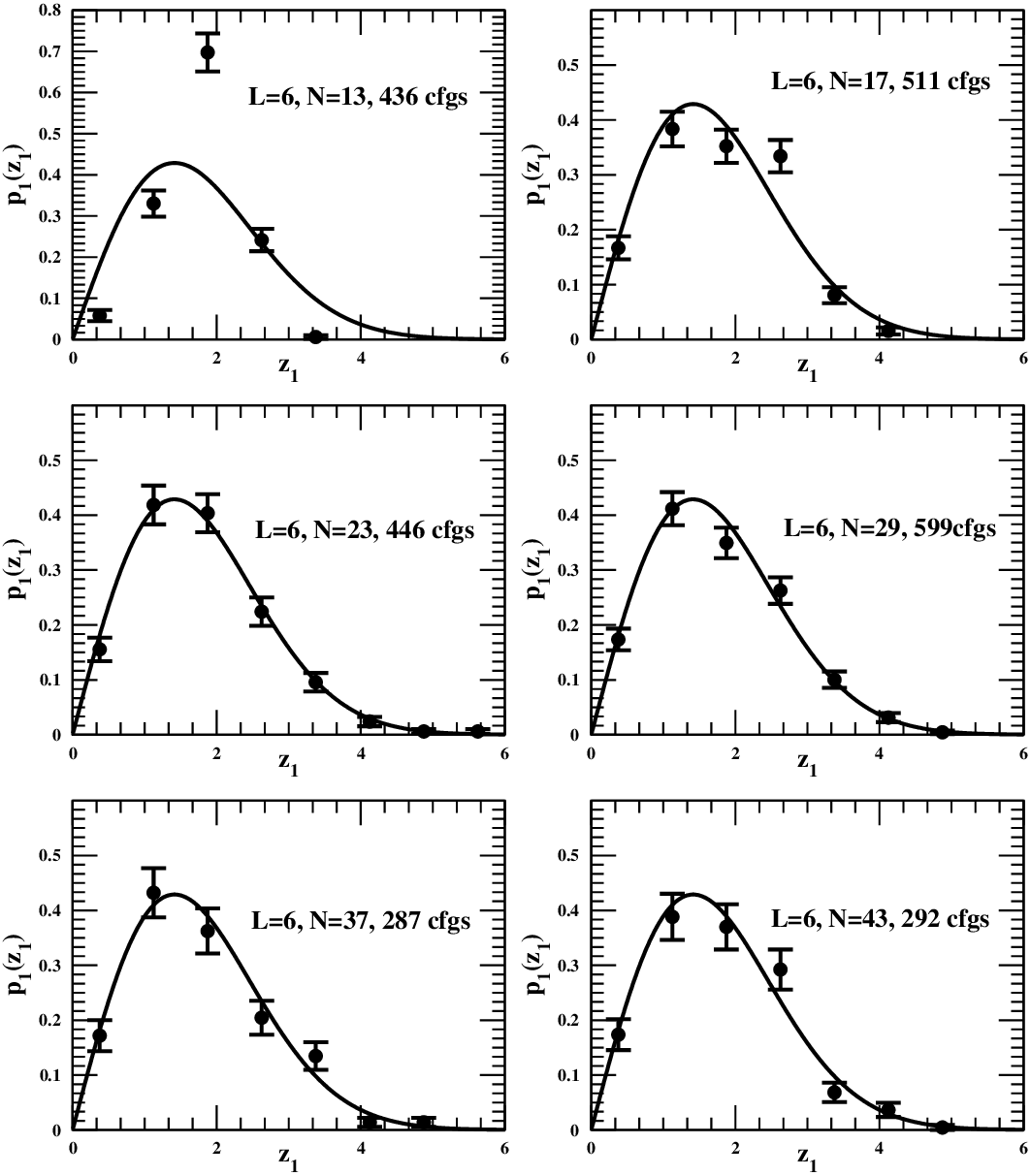}}
\caption{ The distribution of the first eigenvalue scaled by the
corresponding average value
in the zero topological sector for various
$N_c$ at $L=6$ and $b=0.350$ is compared with the prediction from chiral RMT
in four dimensions.
}
\label{hist1l6q0b350}
\end{figure}

\begin{figure}
\epsfxsize = \textwidth
\centerline{\epsfbox{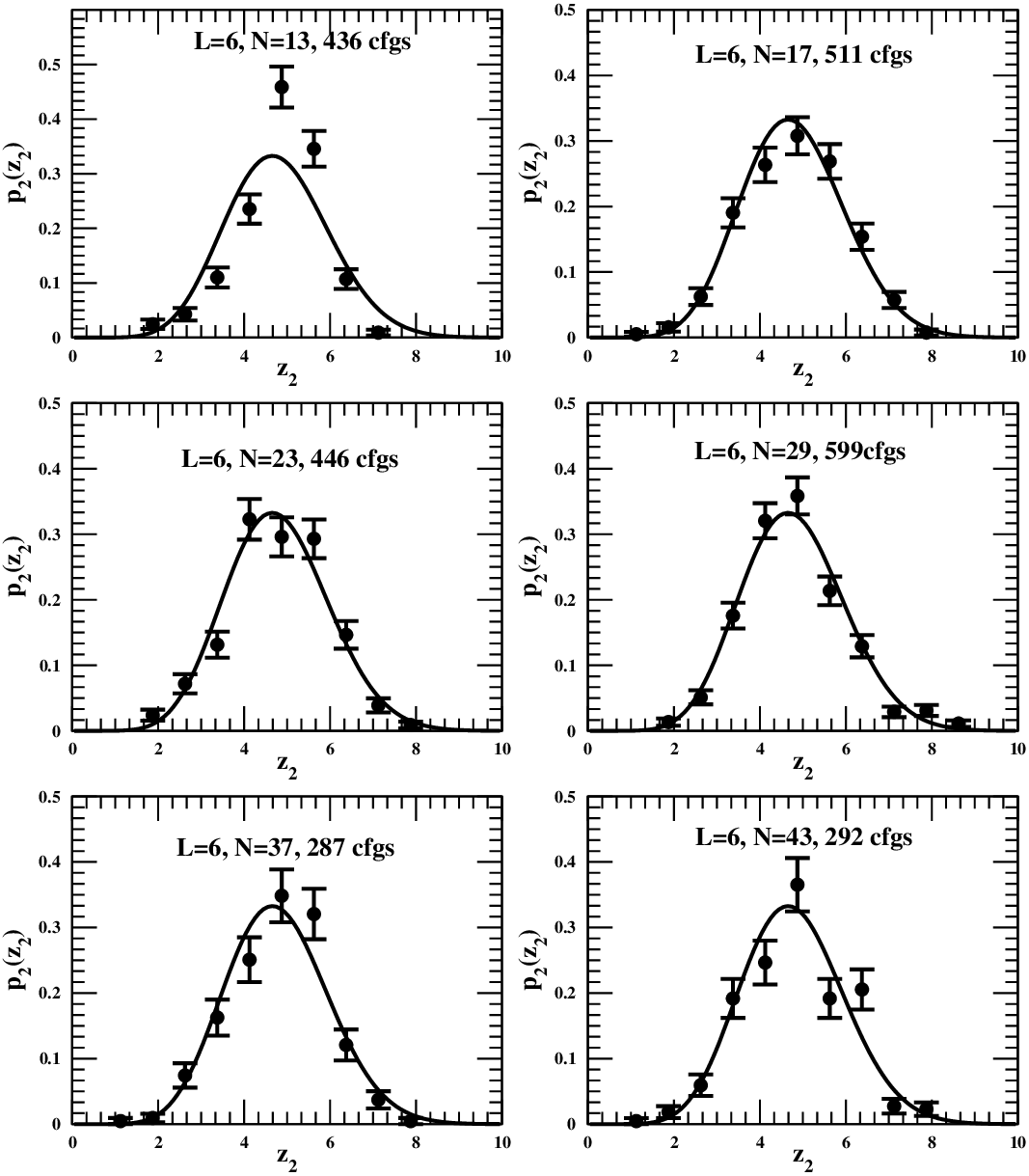}}
\caption{ The distribution of the second eigenvalue scaled by the
corresponding average value
in the zero topological sector for various
$N_c$ at $L=6$ and $b=0.350$ is compared with the prediction from chiral RMT
in four dimensions.
}
\label{hist2l6q0b350}
\end{figure}

We now look at the distributions of the rescaled two eigenvalues
as a function of $N_c$, still at $L=6$ and $b=0.350$ in Fig.~\ref{hist1l6q0b350}
and in Fig.~\ref{hist2l6q0b350}. We see that merely looking at one eigenvalue
in isolation may be misleading
(compare $N_c=17$ to $N_c=23$ for instance). 
A glance at table~\ref{tab1} shows that
the determinations of $\Sigma_i$ differ significantly between $i=1$ and $i=2$ 
for $N_c=13,17$ at $L=6,~b=0.350$, confirming
the indication form the ratio distribution. Again, the scatter in the 
data is consistent with Poisson statistics.

\begin{figure}
\epsfxsize = \textwidth
\centerline{\epsfbox{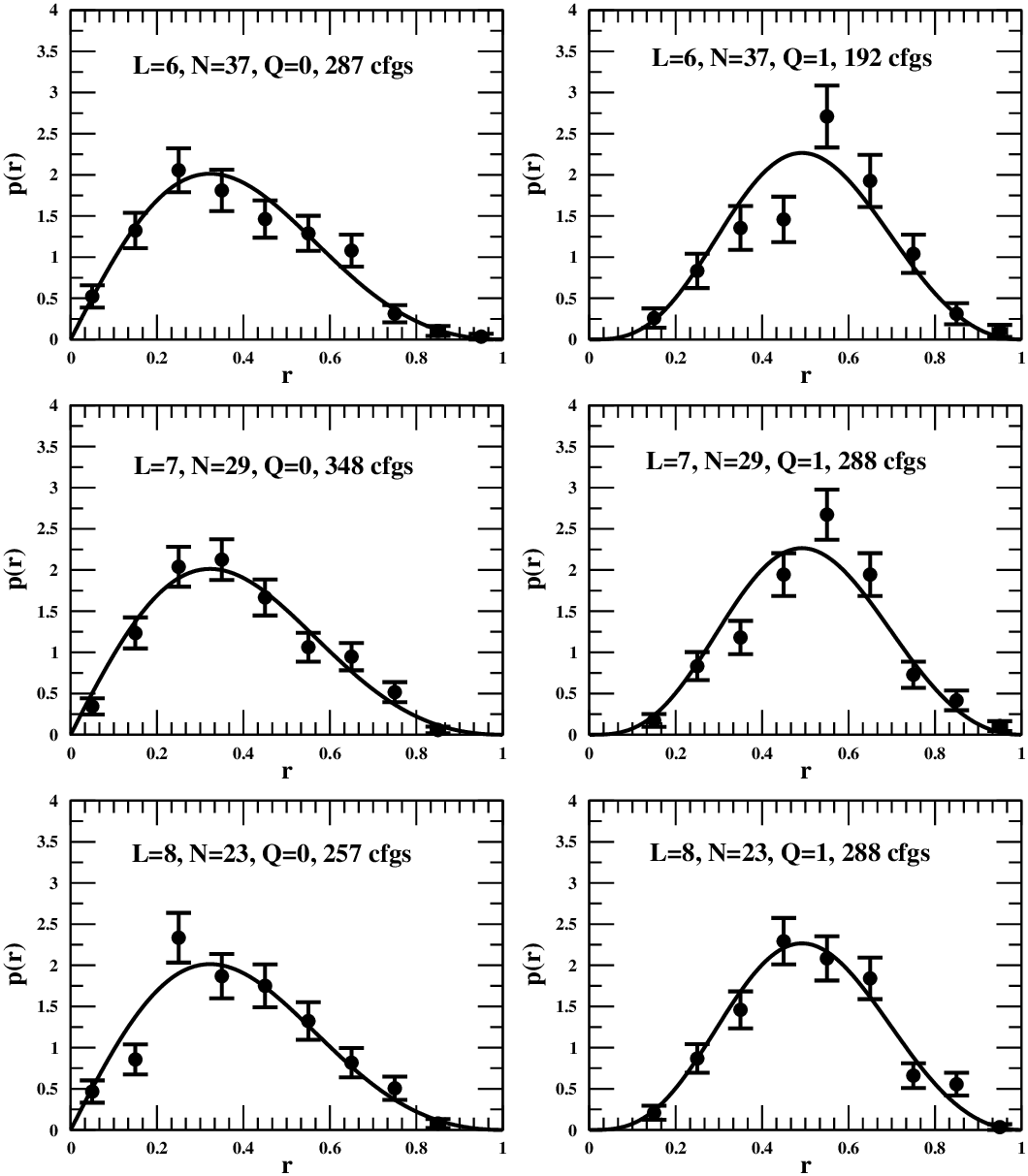}}
\caption{ The distribution of the ratio of the first non-zero eigenvalue to
the second non-zero eigenvalue in the $Q=0,1$ topological sectors for various
$L$ and $N_c$ at $b=0.350$ is compared with the prediction from chiral RMT
in four dimensions.
}
\label{ratiolvqvb350}
\end{figure}

\begin{figure}
\epsfxsize = \textwidth
\centerline{\epsfbox{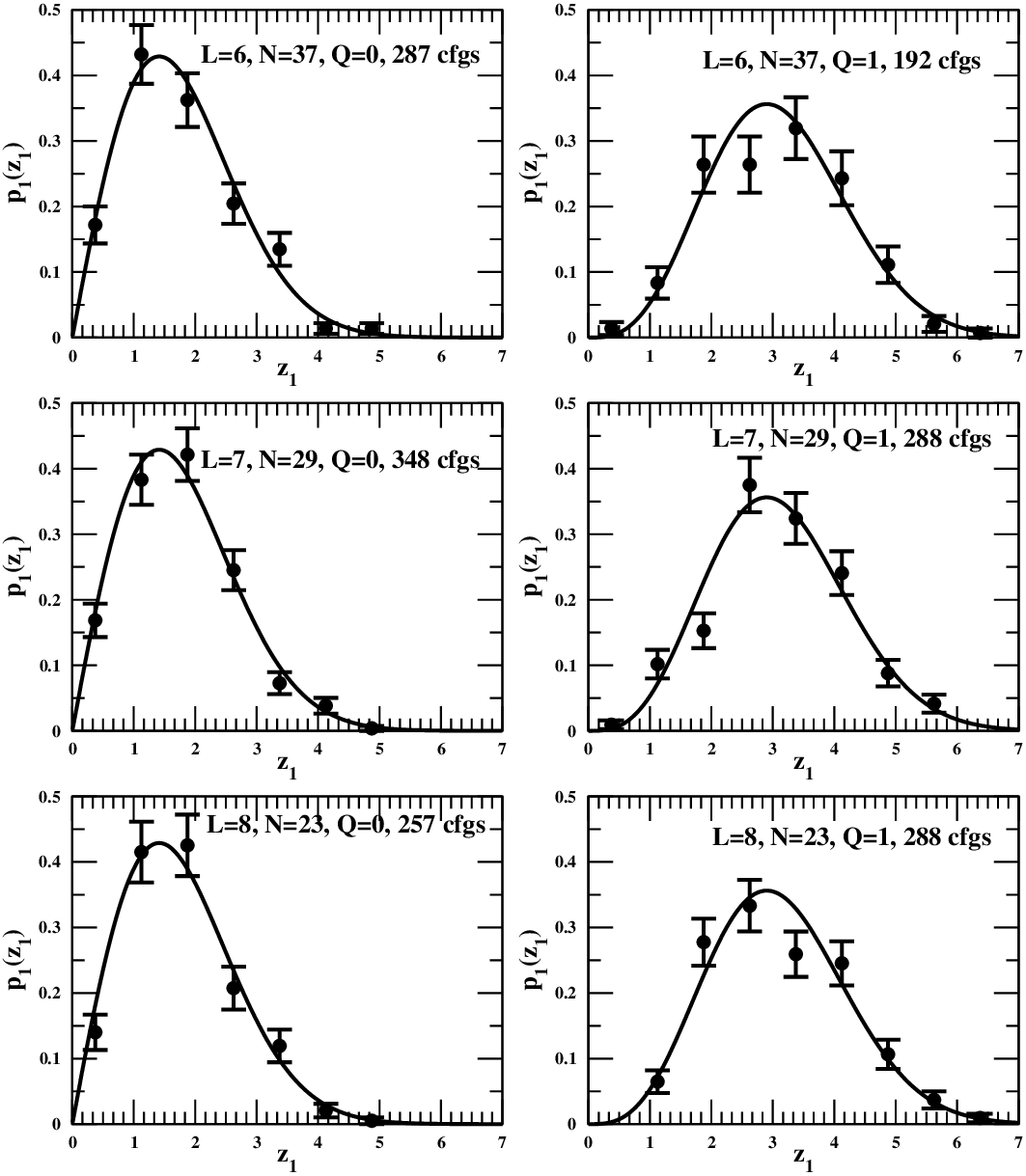}}
\caption{ 
The distribution of the first non-zero eigenvalue scaled by the
corresponding average value
in the $Q=0,1$ topological sectors for various
$L$ and $N_c$ at $b=0.350$ is compared with the prediction from chiral RMT
in four dimensions.
}
\label{hist1lvqvb350}
\end{figure}

\begin{figure}
\epsfxsize = \textwidth
\centerline{\epsfbox{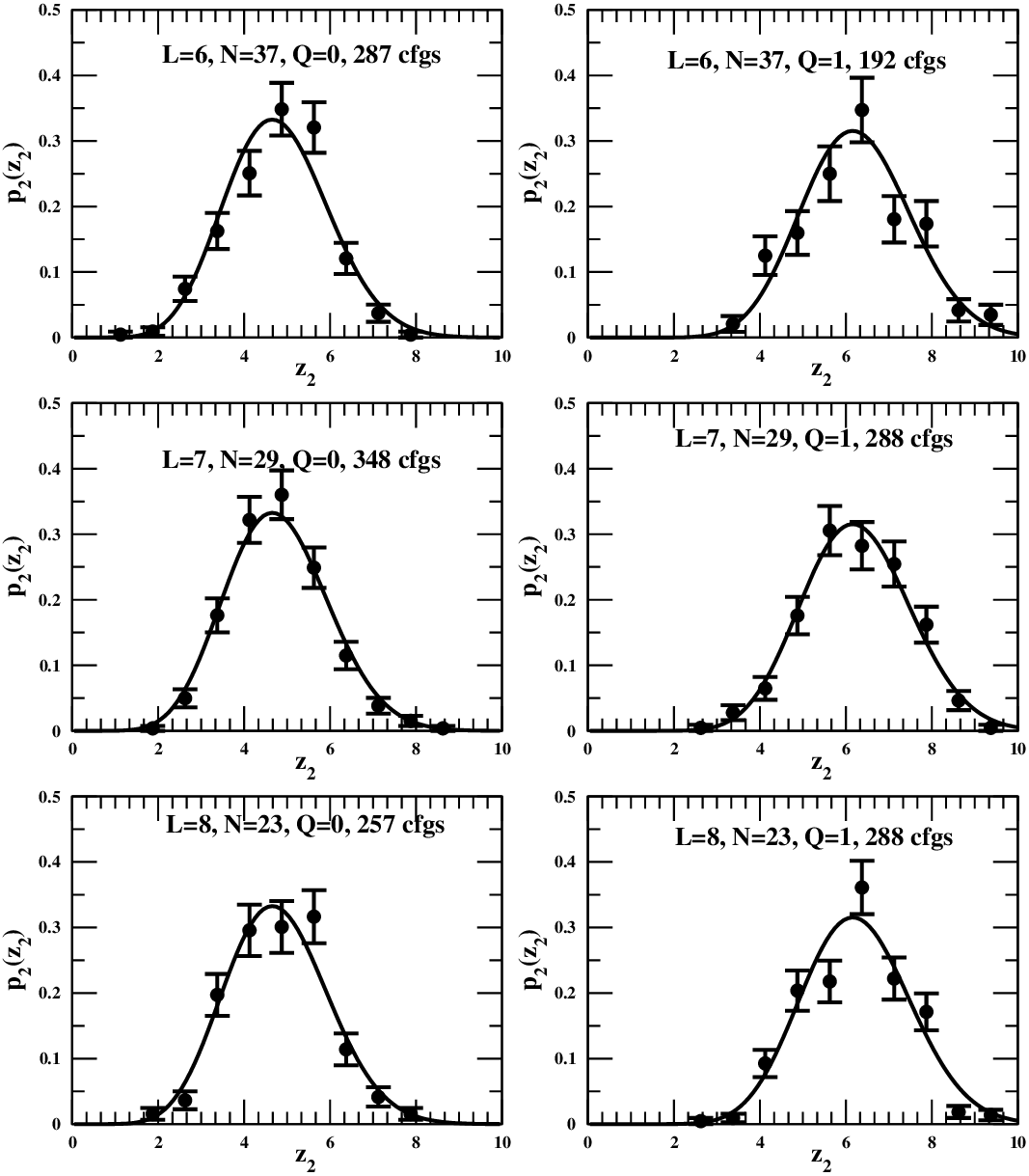}}
\caption{ 
The distribution of the second non-zero eigenvalue scaled by the
corresponding average value
in the $Q=0,1$ topological sectors for various
$L$ and $N_c$ at $b=0.350$ is compared with the prediction from chiral RMT
in four dimensions.
}
\label{hist2lvqvb350}
\end{figure}

We now pursue the question of $L$ dependence by keeping $b=0.350$ and looking for the
infinite $N_c$ limit of the distributions and correctly rescaled eigenvalues for higher
values of $L$. In addition, we wish to ascertain that these distributions depend on
topology in the precise way RMT dictates. Moreover, as table~\ref{tab1} indicates,
the numerical values of the $\Sigma_i$ are topology independent within our statistics. 
The results we would like to look at at this point are collected
in Fig.~\ref{ratiolvqvb350} (for the ratio distributions),
Fig.~\ref{hist1lvqvb350} (for the smallest eigenvalue) 
and Fig.~\ref{hist2lvqvb350} (for the next smallest eigenvalue).
 Our $Q=1$ simulations focused on a value of $N_c$ large enough
that the $Q=0$ case is consistent with RMT already. 
In a marginal case consistency with RMT 
may not hold for $Q=1$ even if it holds for $Q=0$, 
because the two nonzero lowest eigenvalues at
$Q=1$ are repelled by the zero eigenvalue and therefore somewhat higher in magnitude
than in the $Q=0$ case.
The statistical features and fluctuations at $Q=1$ are similar to the ones
at $Q=0$. 

\begin{figure}
\epsfxsize = \textwidth
\centerline{\epsfbox{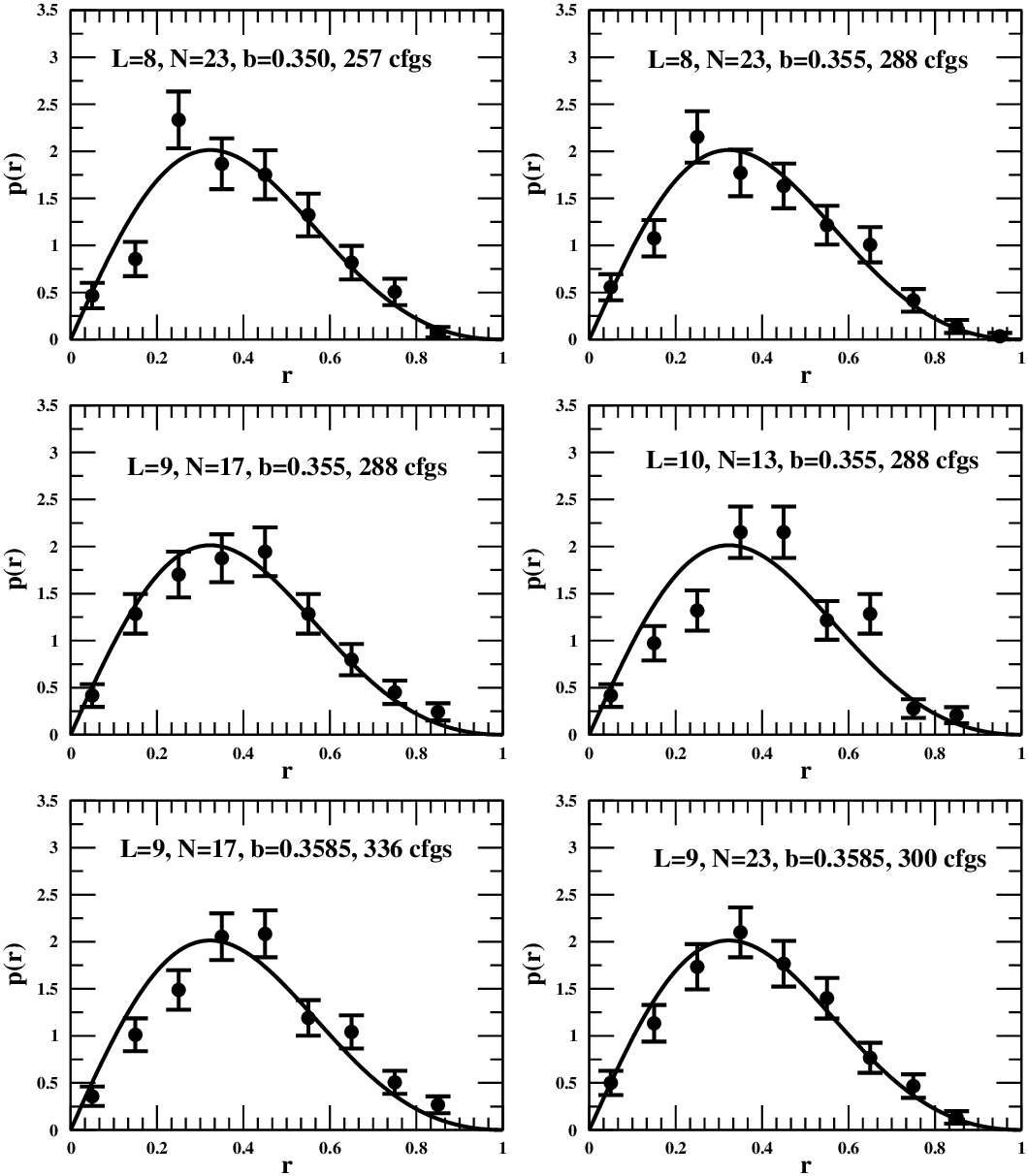}}
\caption{ The distribution of the ratio of the first non-zero eigenvalue to
the second non-zero eigenvalue in the $Q=0$ topological sector for various
$L$, $N_c$ and $b$ is compared with the prediction from chiral RMT
in four dimensions.
}
\label{ratiolvq0bv}
\end{figure}

\begin{figure}
\epsfxsize = \textwidth
\centerline{\epsfbox{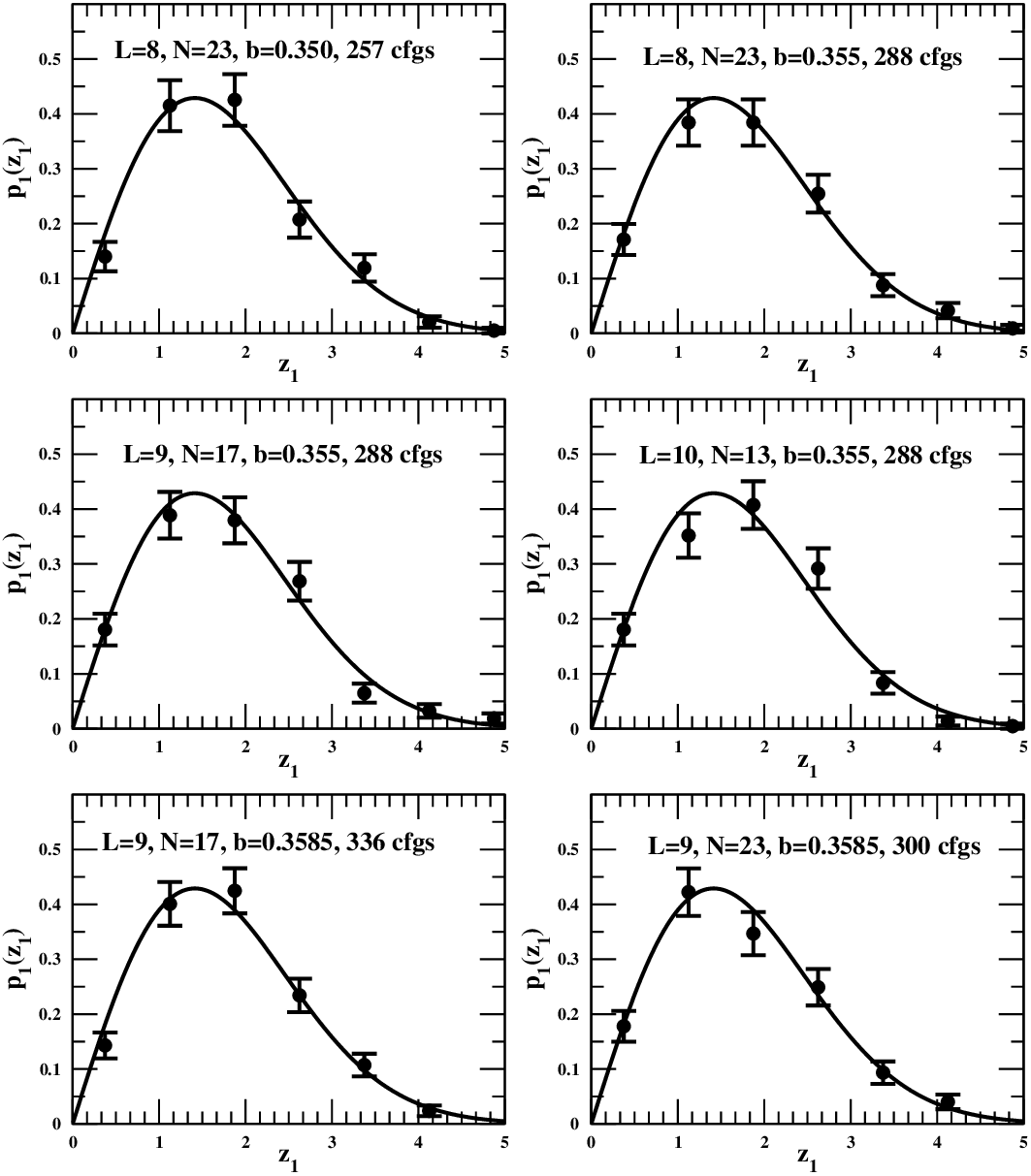}}
\caption{ 
The distribution of the first non-zero eigenvalue scaled by the
corresponding average value
in the $Q=0$ topological sector for various
$L$, $N_c$ and $b$ is compared with the prediction from chiral RMT
in four dimensions.
}
\label{hist1lvq0bv}
\end{figure}

\begin{figure}
\epsfxsize = \textwidth
\centerline{\epsfbox{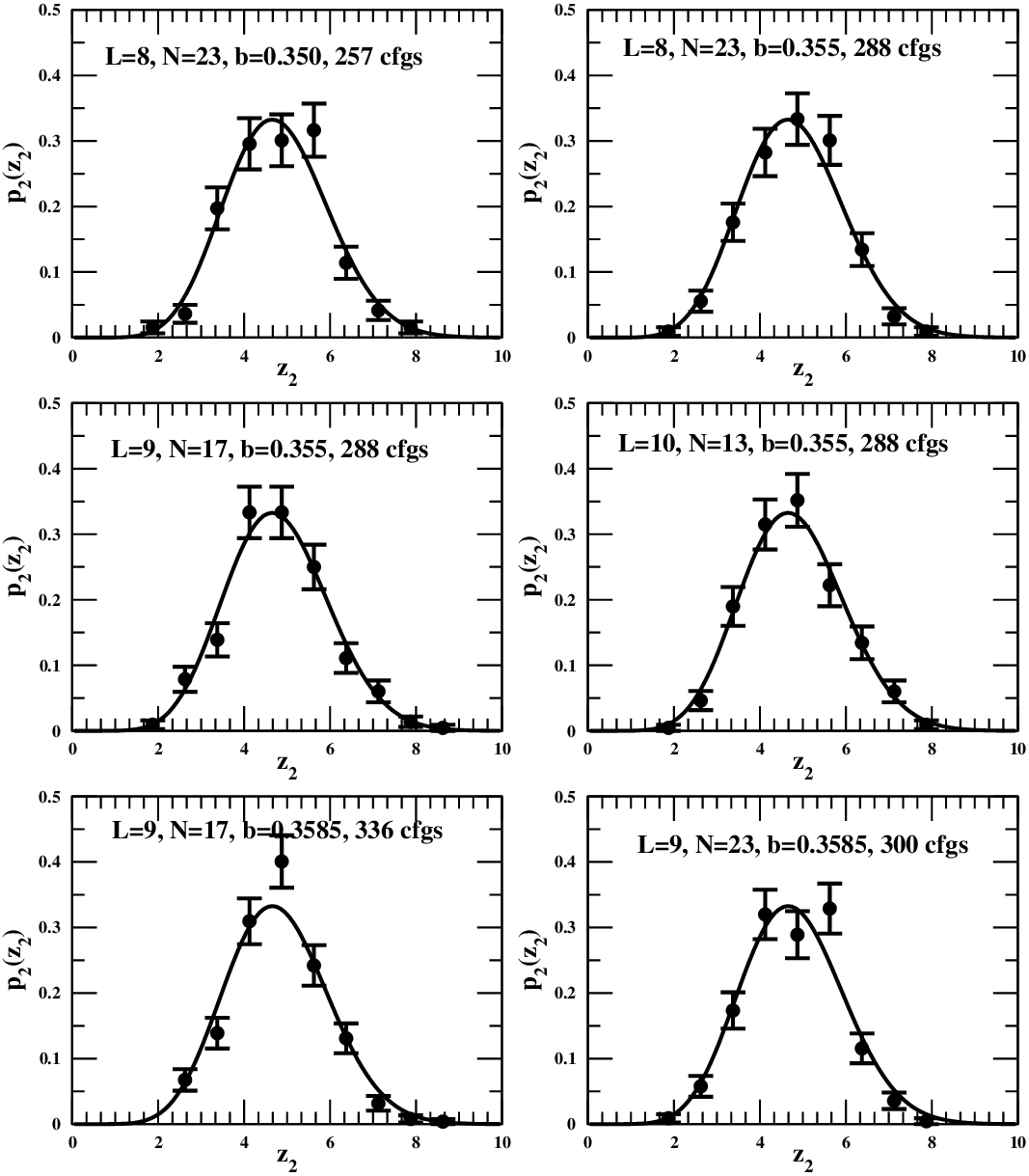}}
\caption{ 
The distribution of the second non-zero eigenvalue scaled by the
corresponding average value
in the $Q=0,1$ topological sectors for various
$L$, $N_c$ and $b$ is compared with the prediction from chiral RMT
in four dimensions.
}
\label{hist2lvq0bv}
\end{figure}

We now pursue other values of $b$, at appropriate volumes and $N_c$ values
where RMT holds and we can focus on the numerical value of $\Sigma$,
ultimately checking consistency with asymptotic freedom scaling.
A sample of results is collected
in Fig.~\ref{ratiolvq0bv} (for the ratio distributions),
Fig.~\ref{hist1lvq0bv} (for the smallest eigenvalue) 
and Fig.~\ref{hist2lvq0bv} (for the next smallest eigenvalue).

\begin{figure}
\epsfxsize = \textwidth
\centerline{\epsfbox{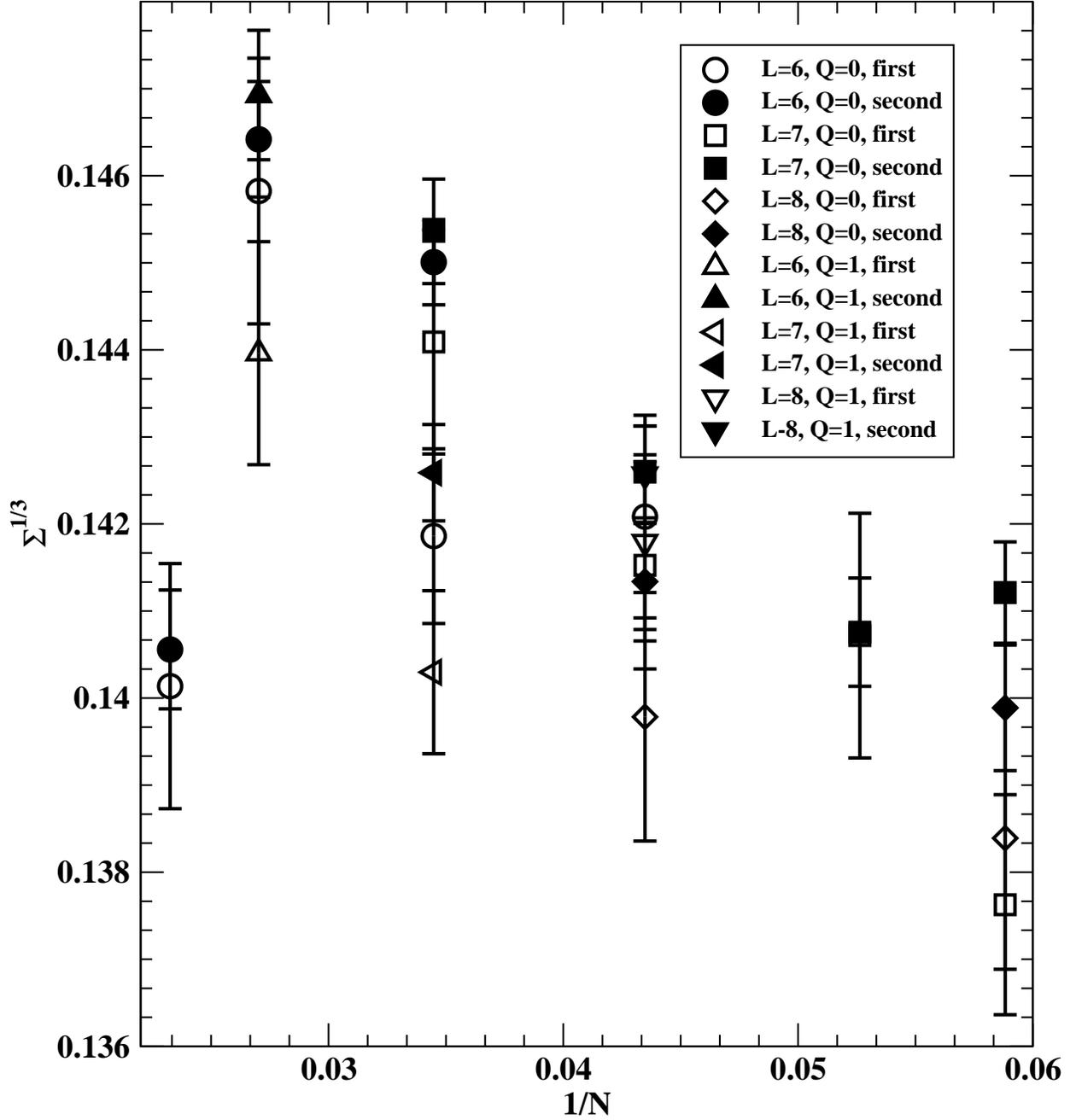}}
\caption{ Estimate of the chiral condensate obtained from the
average of the first and second eigenvalue is plotted against
$\frac{1}{N_c}$ for various $L$ in the $Q=0$ and $Q=1$ topological
sector in four dimensions.
}
\label{sigma}
\end{figure}

We collect our determinations of the chiral condensate at $b=0.350$ 
in Fig.~\ref{sigma}. The data points displayed in this figure show that, within errors,
$\Sigma_1=\Sigma_2$. The errors we displayed for the $\Sigma_i$ are 
obtained by assuming that the law of large numbers holds and 
the average of the individual eigenvalue is 
Gaussianly distributed. Our statistics are somewhat meager and 
therefore it is safer to view our numbers as reliable to within two or three
standard deviations. 

To get some feeling for the physical order of magnitudes we are getting we consider the
$y$-axis in Fig.~\ref{sigma}; we see that $\Sigma^{\frac{1}{3}}$ has  a range of
$0.142\pm 0.006$. Ignoring the scalar density renormalization factor we can get
an approximate number for the continuum value of the condensate. 
\be
\Biggl[\frac{\Sigma(b=0.35)}{2|M|}\Biggr]^{1/3} = 0.098\pm 0.004
\ee
where $M$ is the Wilson mass that enters the overlap Dirac operator.
We have set $M=-1.5$ in our calculations.
Taking into account that the lattice critical size $L_c(b)\approx 6$ at 
$b=0.350$ and assuming that $l_cT_c=1$, we get
$\Sigma^{1/3}_{R,{\rm cont}}(b=0.35)=0.588T_c$. 
Taking $T_c\approx 0.6\sqrt{\sigma}\approx 264$MeV~\cite{teper} gives
$\Sigma^{1/3}_{R,{\rm cont}}\approx 155$MeV. If we now assume that $N_c=3$ is large enough
to ignore $\frac{1}{N_c}$ corrections, $\langle\bar\psi\psi\rangle_{N_c=3}\approx (224{\rm MeV})^3$. 
In spite of the 
roughness of our assumptions, the number comes out close to experiment, 
but one should not make too much of this, because the renormalization factor
can easily change the scale by as much as 25\%. 

One can try to get a rough feel for the latter effect by using
tadpole improved one loop estimates~\cite{tadpole} for $Z_S$ in the $\overline{MS}$
scheme~\cite{horsley}:
\be
Z_S^{\rm {\overline {MS}}} (\mu ) =1.40\{1-\frac{0.006}{b} [ 6\log (a\mu ) -1.33] \}
\ee
The prefactor $1.40$ is $\frac{|M|}{|M|_{\rm tad}}$, where the 
tadpole-improved~\cite{tadpole} Wilson mass is given by
\be
-M_{\rm tad}=|M|_{\rm tad} = \frac{ |M| - 4 (1-u_0)}{u_0}
\ee
$u_0$ is given by the fourth root of the average plaquette, a number 
quite close to unity.
The entire one loop correction is numerically negligible for $\mu \sim 2~{\rm GeV}$, so
$Z_S^{\rm {\overline {MS}}} (2~{\rm GeV}) =1.40$. 
This increases the scale by about 12\%, to $\frac{1}{N_c} \langle\bar\psi\psi\rangle\approx 
(174{\rm MeV})^3$. For QCD this translates into $\langle\bar\psi\psi\rangle_{\rm QCD} \approx 
(251{\rm MeV})^3$, close to results obtained in quenched lattice QCD~\cite{beci}. 
Obviously, before a scale can be reliably quoted 
one would need to determine $Z_S$ outside perturbation theory, by more numerical work.

\subsection {Extracting the chiral condensate from mass dependence}

The conventional method to estimate the chiral condensate is to obtain
a stochastic estimate of the fermion bilinear $\langle \bar\psi \psi \rangle (\mu)$ 
as a function of fermion
mass $\mu$ and perform an extrapolation to zero $\mu$. Since, strictly speaking, we
define the large $N_c$ limit of the massless theory by first keeping $\mu$ finite
while taking $N_c$ to infinity, and only subsequently 
taking $\mu$ to zero, the conventional
method is the most conservative way to go about 
establishing S$\chi$SB and estimate the order parameter
$\langle \bar\psi \psi \rangle (0)$. However, it is 
not practical to make $N_c$ large enough
to disentangle the effect we are after from finite $N_c$ effects. 
As we shall see, the RMT
route is much better in practice at determining the infinite 
$N_c$ behavior from finite $N_c$ results.

We evaluate $\langle b| {A^\dagger}^{-1}(\mu) A^{-1}(\mu)
| b \rangle $ for a chiral ``source'' vector $|b>$, 
randomly drawn from a Gaussian distribution, and obtain
a stochastic estimate of the trace we are interested in:
\be
F_1(\mu) = Tr {A^\dagger}^{-1}(\mu) A^{-1}(\mu)
\ee
We then estimate the condensate from
\be
\Sigma^{1/3} = \lim_{\mu\rightarrow 0} F_1^{1/3}(\mu).
\ee

\begin{figure}
\epsfxsize = \textwidth
\centerline{\epsfbox{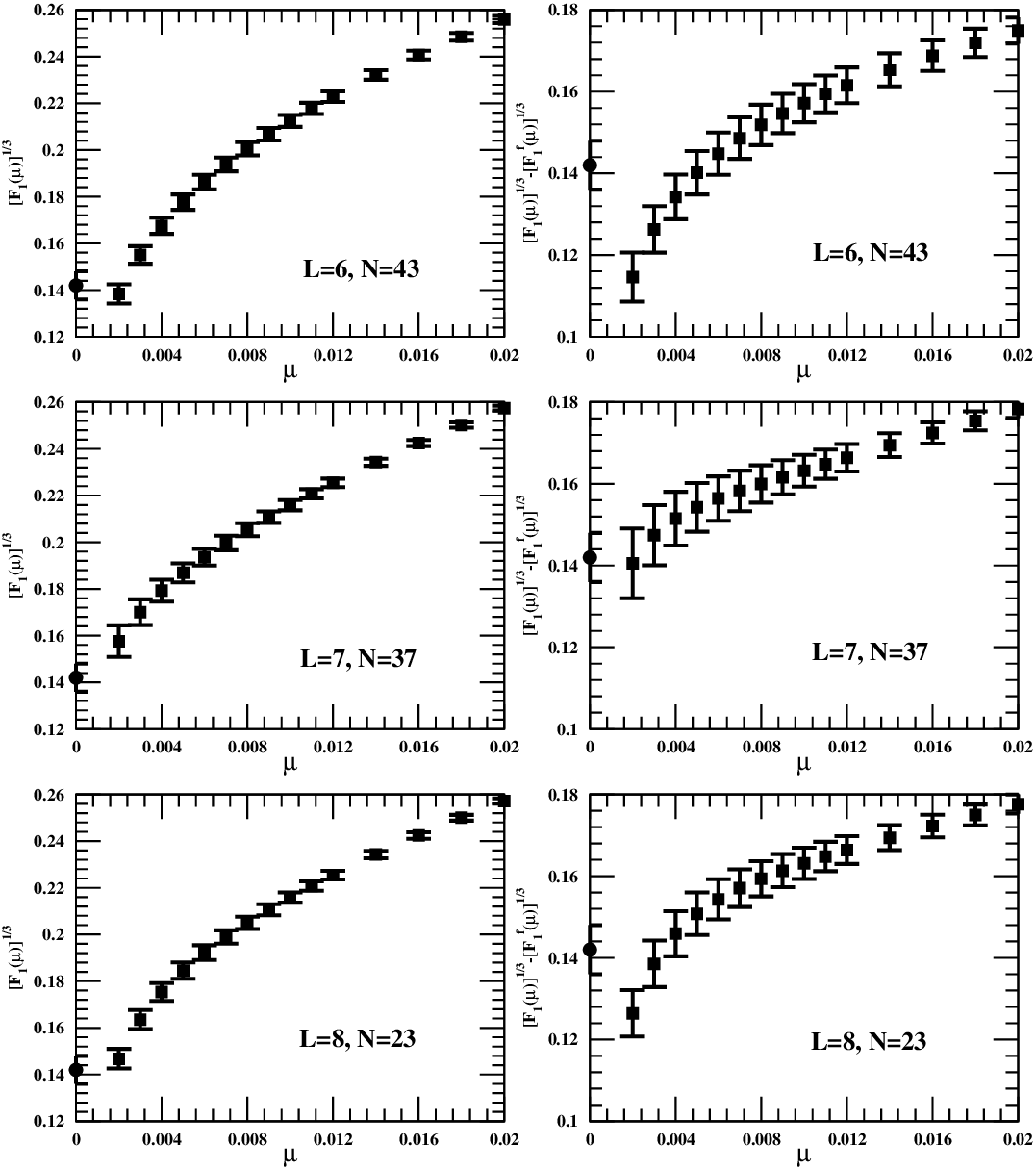}}
\caption{ The three panels on the left side show the stochastic
estimate of $F_1^{1/3}(\mu)$ in four dimensions for three different
combinations of $L$ and $N_c$. 
The three panels on the right side show the same estimate after
the subtraction of the tadpole corrected free field contribution.
The solid circle with errorbars is
the estimate for the chiral condensate from chiral RMT.
}
\label{pbp4d}
\end{figure}

Using one random source per configuration, we computed 
$F_1(\mu)$ on $48$ different thermalized gauge background 
configurations at three different values of
$(L,N_c)$. These simulations were all done at our 
chosen coupling, $b=0.350$.
The results are plotted in the three left panels
of Fig.~\ref{pbp4d} and compared with
the estimate of $\Sigma^{1/3}$ from the previous subsection.
The stochastic estimate seems to be consistent with the
estimate from chiral random matrix theory but the range covered
on the $y$-axis is too large to make this claim convincing. 

A more careful examination reveals that much of $F_1(\mu)$ comes from
a free field contribution and one could subtract that free field contribution
by an explicit computation, without affecting the $\mu\to 0$ limit so long
as there is S$\chi$SB. The coupling is not weak, 
so one must use the ``right'' free field case: 
The natural choice is to employ the tadpole-improved Wilson mass
in $H_w$, when computing the free fermion contribution.
Denoting the free field result by $F^f_1(\mu)$ we carried out the subtraction and the
result is plotted in the three right panels of
Fig.~\ref{pbp4d}. We see that indeed $F_1(\mu)$
is dominated by the free fermion contribution and that the subtracted
quantity still is consistent with the chiral random matrix theory prediction
for $\Sigma^{1/3}$.

\begin{figure}
\epsfxsize = \textwidth
\centerline{\epsfbox{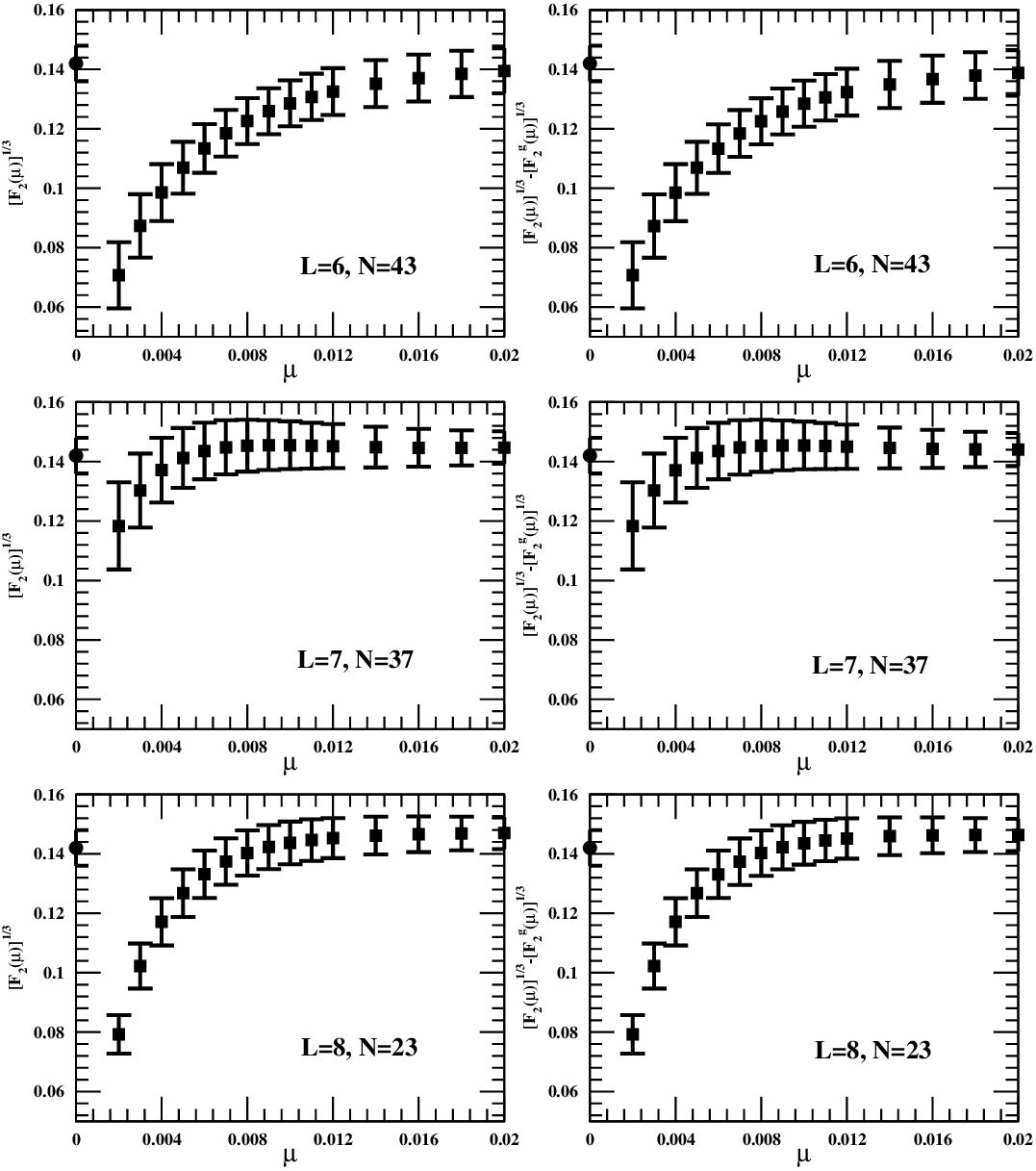}}
\caption{ The three panels on the left side show the stochastic
estimate of $F_2^{1/3}(\mu)$ in four dimensions for three different
combinations of $L$ and $N_c$. 
The three panels on the right side show the same estimate after
the subtraction of the tadpole corrected free field contribution.
The solid circle with errorbars is
the estimate for the chiral condensate from chiral RMT.
}
\label{chi4d}
\end{figure}

As explained in Section~\ref{overview}, 
one can improve the situation by defining a quantity that has
less of a divergence in the ultraviolet:
\be
F_2(\mu) = 2\mu^3 Tr \Bigl [ {A^\dagger}^{-1}(\mu) A^{-1}(\mu) \Bigr ]^2
\ee
It is easy to see that
\be
\lim_{\mu\rightarrow 0} F^{1/3}_2(\mu) = \Sigma^{1/3}
\ee
This is a particular attractive alternative, since a 
stochastic estimate of $F_2(\mu)$ needs no
more numerical computation than what is needed for $F_1(\mu)$. 
We note that
\be
F_2(\mu) = -2\mu^3 \frac{d}{d\mu^2} \left ( \frac{F_1(\mu)}{\mu} \right )
\ee
and therefore the ultraviolet divergent term, which is linear in $\mu$  
and additively contributes to $F_1(\mu)$, does not contribute to $F_2(\mu)$. 
The first sub-leading term in $F_2(\mu)$
is proportional to $\mu^3$ and therefore 
we expect a better estimate of $\Sigma^{1/3}$ from 
$F_2(\mu)$. The results for $F_2(\mu)$ are plotted in the three left panels
of Fig.~\ref{chi4d}; these results are obtained simultaneously 
with those for $F_1 (\mu)$, at no additional
numerical cost. 
Again, the results seem consistent with the prediction obtained using
chiral random matrix theory in the previous subsection.
Similarly to $F_1(\mu)$, there is a ``free field'' contribution to
$F_2(\mu)$ and we subtracted this ``free part'', 
$F_2^f(\mu)$ (with tadpole improved parameters), by a direct computation.
The resulting subtracted quantity is plotted in 
the three right panels of Fig.~\ref{chi4d}
and one can see that this time the subtraction had no discernible effect.
As expected, the subtraction of the free field expression mainly affects
the ultraviolet contribution. 
Consistency with the RMT results is better for $F_2 (\mu )$ than for $F_1 (\mu)$,
but the RMT approach still appears significantly superior in practice. 

\subsection {Approximate scaling}

As explained, the coarsest evidence for scaling should come by simply observing that
the quantity $s(b)\equiv L_c (b) \left (\frac{\Sigma (b)}{2|M|}\right )^{\frac{1}{3}}$ 
stays constant as $b$ is varied. The lattice critical size $L_c (b)$ has been determined
numerically in ~\cite{cek4} to be given by:
\begin{eqnarray}
L_c (b) &=& (0.260\pm 0.015) \left ( \frac{11}{48\pi^2 b_I} \right )^{51/121} e^{24\pi^2 b_I /11}\cr
b_I &\equiv& b \frac {b^2 -0.58964b+0.08467}{b^2-0.50227b+0.05479}
\end{eqnarray}
This formula uses tadpole improvement and provides an interpolation valid for $6\le L_c(b) \le 10$.
For a given $b$ the relevant calculations are carried out at $L \ge L_c(b)$. 

Looking at the table we find $s(0.350)=0.59(6)$, $s(0.355)=0.58(6)$ and $s(0.3585)=0.58(6)$.
Even a limited investigation on a very coarse lattice gives a consistent number: $s(0.346)=0.55(5)$.
If we include the factor $Z_S^{\rm {\overline {MS}}} (2~{\rm GeV})=1.40$, 
we obtain a continuum statement at infinite $N_c$:
\be
\frac{l_c^3}{N_c} \langle\bar\psi\psi\rangle^{\rm {\overline {MS}}}( 2~{\rm GeV})\approx (0.65)^3
\ee

\section {Numerical results in two dimensions}

Two dimensional large $N_c$ QCD~\cite{hooft2d} can be used as a soluble example for the
phenomenon of S$\chi$SB at infinite $N_c$.
In the absence of exact zero modes, 
just as in four dimensions, we expect RMT to work; moreover,
it is the same RMT model that applies. Also, one can define in entirely
analogous ways the functions $F_1 , F_2$. The major advantages are: simulations
are relatively very fast, enabling one to compute the entire spectrum of $A$ and, in addition, 
the continuum value of $\Sigma$ is known from the analytical solution of 't Hooft. 

For completeness, we start with a brief 
derivation of the exact value of the chiral condensate in 
large $N_c$ QCD~\cite{hooft2d,2d}.
Taking care of the wavefunction normalization, we have
\be
Tr {A^\dagger}^{-1}(\mu) A^{-1}(\mu) = 4M^2 \frac{r_0^2}{\mu_0^2} \label{chi}
\ee
in the limit of small quark mass,  
where only the first term in the infinite sum over poles is expected to contribute 
(for this to be true the infinite sum over poles needs to be regulated to eliminate
ultraviolet divergences - because the underlying model is renormalizable, 
it does not matter how exactly this regularization is done) 
and $r_0$ is the corresponding residue. $\mu_0^2$ is the ``pion'' 
mass squared.
The quark mass is related to $\mu$ by $m_q=2|M|\mu$.
The quark mass in units of the gauge coupling is given by a 
dimensionless parameter $\gamma$:
\be
m_q = \frac{\sqrt{\gamma}}{\sqrt{2\pi b}} \label{mass}
\ee
Using (\ref{chi}) and (\ref{mass}), we get
\be
\Sigma = \frac{2|M|\sqrt{\gamma}}{\sqrt{2\pi b}}\frac{r_0^2}{\mu_0^2}
\ee
The lowest pseudoscalar mass is given by
\be
\mu_0^2 = \frac{2\pi\sqrt{\gamma}}{\sqrt{3}}
\ee
and its residue is given by
\be
r_0^2 = \frac{\pi}{3}
\ee
Putting together all the information, we get
\be
\Sigma = \frac{|M|}{\sqrt{6\pi b}}
\ee
We will use $M=-1$ and $b=1$ for our numerical analysis. 
Previous experience~\cite{2d} tells us that $b=1$ is sufficiently
large to expect the effects of finite lattice spacing to be
relatively small, an expectation confirmed by our findings here too. 
This provides a numerical value for the lattice 
condensate: $\Sigma=\frac{1}{\sqrt{6\pi}}=0.2303$.

The numerical computation using overlap fermions in two dimensions
was done in a manner similar to the one in four dimensions. However, we 
performed a full diagonalization of the overlap Dirac operator
in two dimensions, eliminating this potential source of statistical 
noise (in practice, the fluctuations associated with the stochastic
nature of the trace evaluation are substantially smaller than those
associated with varying the gauge field background). 
Thus, we did not resort to a
stochastic estimate of $F_1(\mu)$ and $F_2(\mu)$ in two dimensions and, as we shall see presently, 
obtained data very similar to four dimensions, where we are forced to use a stochastic
estimator for the trace. We generated a
total of $1024$ configurations at a fixed $L$ and $N_c$. We worked
with the following $(L,N_c)$ pairs: $(4,57)$, $(5,47)$, $(6,37)$
and $(7,31)$. All configurations were gauge field backgrounds with no
exact fermion zero modes.

\begin{figure}
\epsfxsize = \textwidth
\centerline{\epsfbox{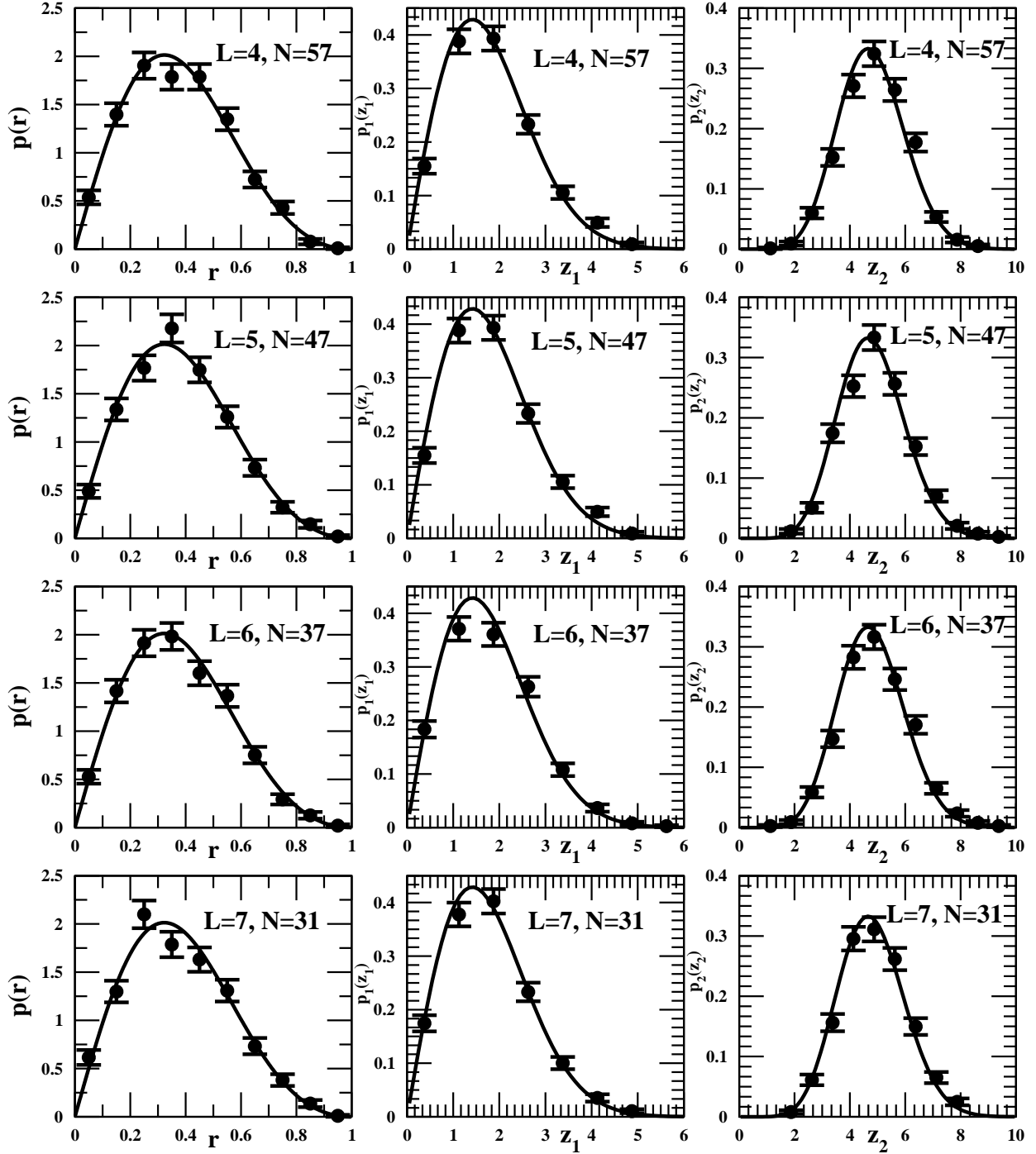}}
\caption{The distribution of the ratio of the first eigenvalue to
the second eigenvalue in the zero topological sector for various
$L$ and $N_c$ is compared with the prediction from chiral RMT in
two dimensions in the four left panels.
The distribution of the first and second eigenvalue, scaled by the
corresponding average value
in the zero topological sector, for various
$L$ and $N_c$, is compared with the prediction from chiral RMT
in two dimensions in the four middle and right panels respectively.}
\label{rmt2d}
\end{figure}

Like in four dimensions, we start with a plot of the distribution
of the ratio of the first eigenvalue to the second eigenvalue
and compare it with chiral RMT. This is shown in the
four left panels in Fig.~\ref{rmt2d}.
As in four dimensions, the agreement is fairly good and one sees
large scatter at values of $r$ where $p(r)$ is large.
The four middle panels and the four right panels in
Fig.~\ref{rmt2d} show the
comparison between data and chiral RMT for the two lowest eigenvalues.
In relating the chiral RMT variable $z$ to the eigenvalue $\lambda$ of the
overlap Dirac operator via $z=\lambda \Sigma N_c L^2$, 
we used the known result for the condensate, namely, 
$\Sigma=0.2303$. Our main point is that
we know in this case that S$\chi$SB occurs and 
the plots still have the same general
structure as in four dimensions where 
we wanted to determine whether S$\chi$SB takes place.

\begin{figure}
\epsfxsize = \textwidth
\centerline{\epsfbox{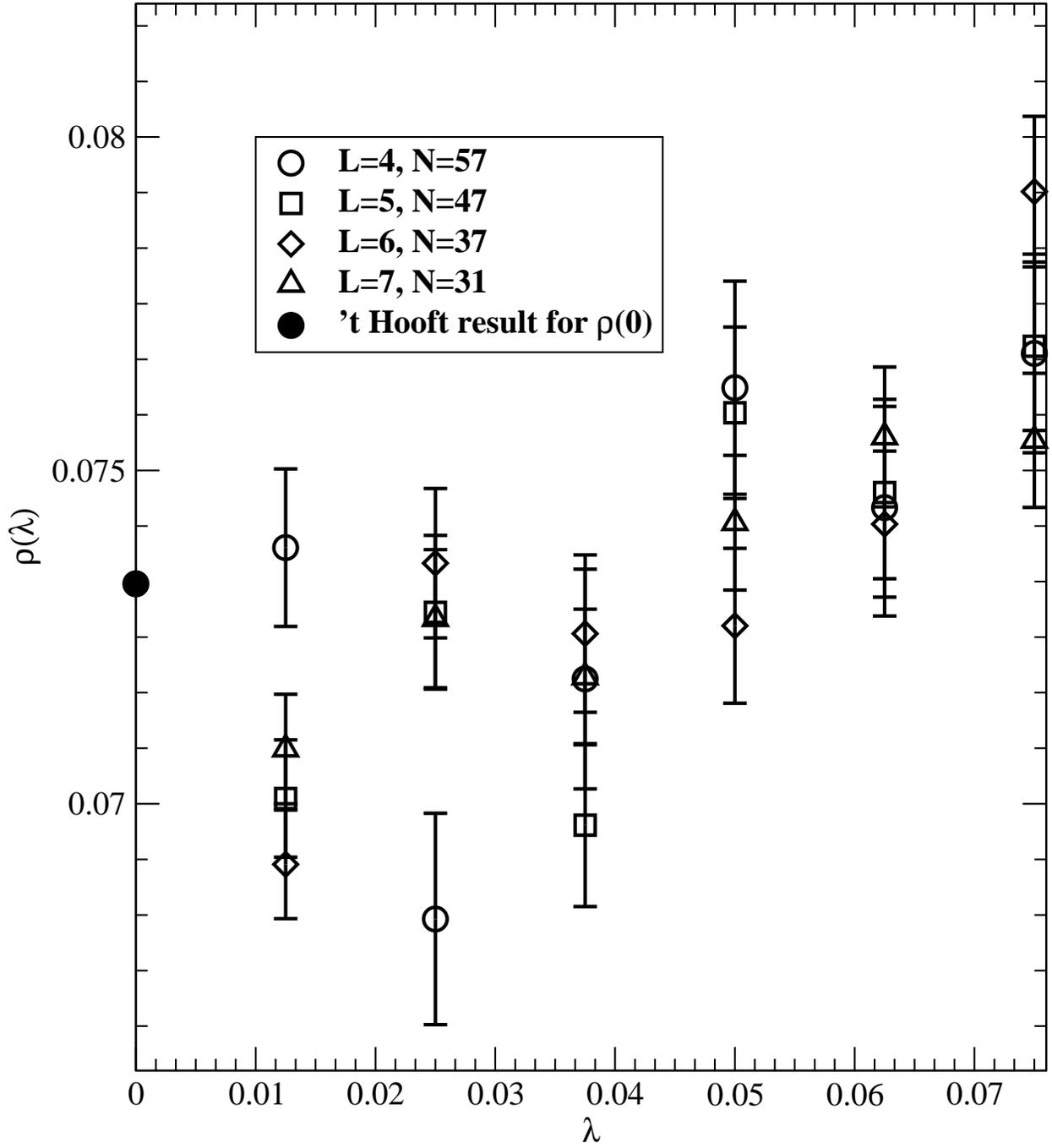}}
\caption{ 
A plot of the spectral density $\rho(\lambda)$ in two dimensions. 
}
\label{rhob2d}
\end{figure}

Since we have computed all the eigenvalues of the overlap Dirac operator,
we can also obtain a plot of the spectral density, $\rho(\lambda)$. This
is impossible to do in four dimensions, except, maybe, for small values of $N_c$.
The result is in Fig.~\ref{rhob2d},  
where we show the distribution only close to $\lambda=0$.
Getting the eigenvalue density even in this limited range would be very costly in four
dimensions. As is well known~\cite{bc}, $\rho(0)=\Sigma/\pi$
and we see that the theoretically expected number, $0.0733$, is correctly reproduced, 
with an accuracy and consistency similar to that obtained in four dimensions, using RMT.

\begin{figure}
\epsfxsize = \textwidth
\centerline{\epsfbox{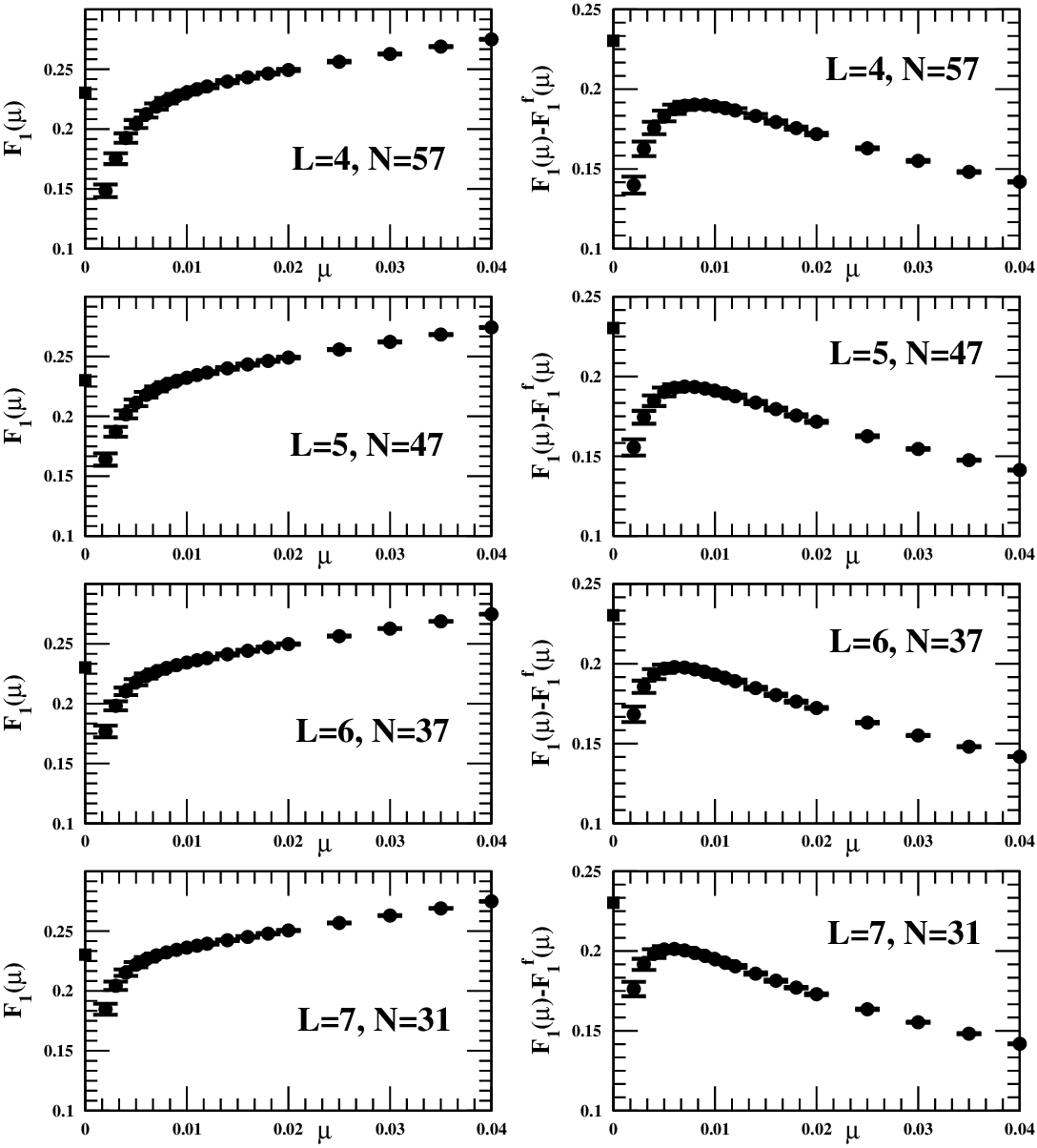}}
\caption{ 
The four panels on the left along with the four panels
on the right show plots of
$F_1(\mu)$ in two dimensions before and after the subtraction
of the free field contribution.
The solid square present in all
the panels is the exact continuum result for the chiral condensate.
}
\label{pbp2d}
\end{figure}

\begin{figure}
\epsfxsize = \textwidth
\centerline{\epsfbox{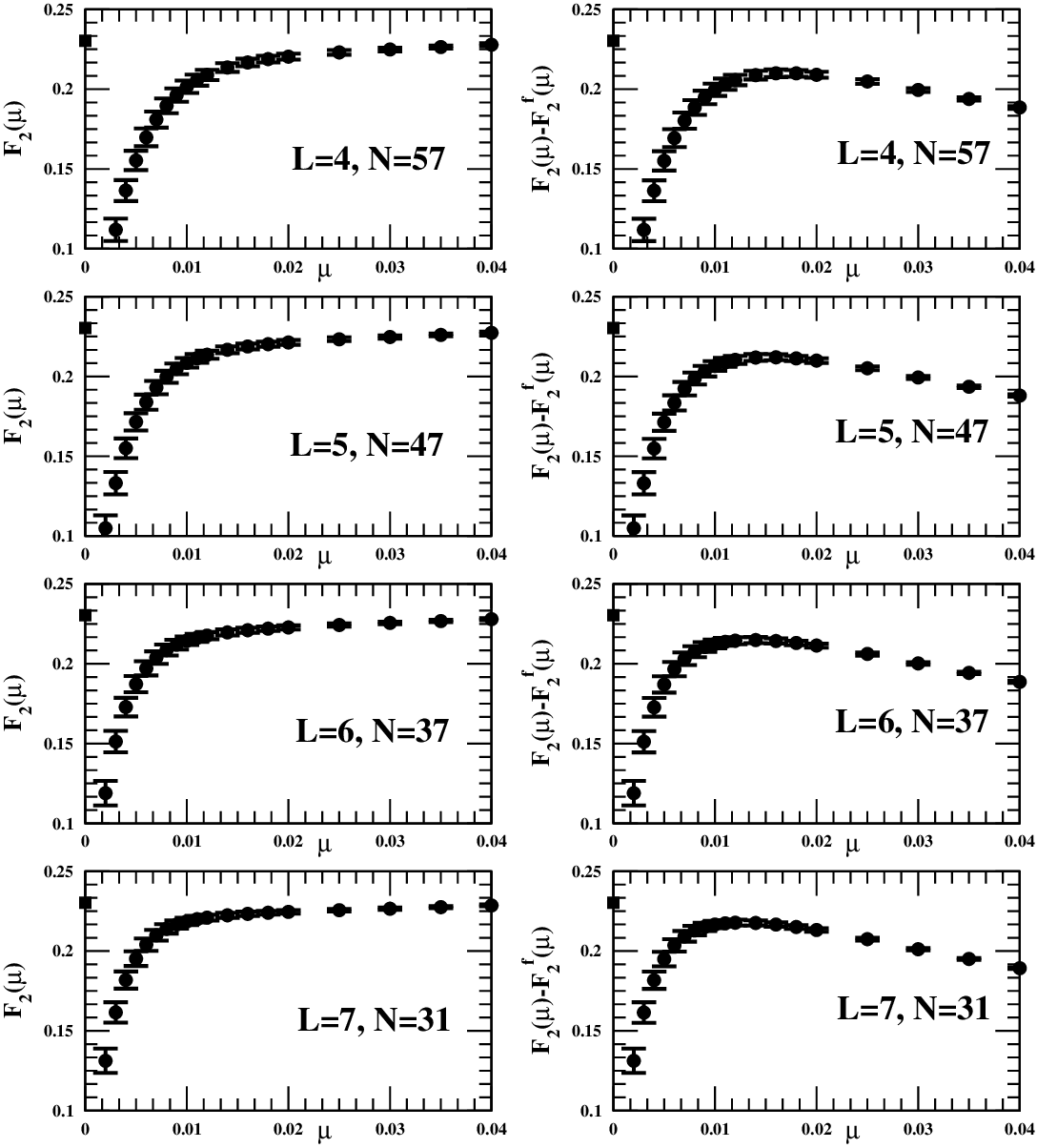}}
\caption{ 
The four panels on the left along with the four panels
on the right show plots of
$F_2(\mu)$ in two dimensions before and after the subtraction
of the free field contribution.
The solid square present in all
the panels is the exact continuum result for the chiral condensate.
}
\label{chi2d}
\end{figure}

We can try to see if one can also get the chiral condensate from a
direct extrapolation of $F_1(\mu)$ and $F_2(\mu)$. Plots of $F_1(\mu)$
and $F_2(\mu)$ along with their corresponding result after the
subtraction of the tadpole improved free field contribution are shown
in the various panels of Fig.~\ref{pbp2d} and Fig.~\ref{chi2d}. 
We see that these more
direct observables indeed are poorer practical indicators for
S$\chi$SB and give less accurate estimates of the condensate if
S$\chi$SB is assumed. The overall structure of the data is similar to
four dimensions, indicating that nothing is wrong with our
interpretation of the four dimensional results.

\newpage

\section {Summary}

Our evidence supports the basic hypothesis of this paper: that, in the planar limit,
continuum $SU(N_c)$ gauge theory, defined on an Euclidean four 
dimensional torus of side $l$, breaks
chiral symmetry spontaneously so long as $l>l_c$, where $l_c$ is a physical length
of order $1~{\it fermi}$ in QCD units.

Some may find it surprising that a continuous symmetry is spontaneously 
broken in a finite volume.
Maybe the following rather trivial observations would convince the 
skeptics that this is not that 
unexpected. Suppose we consider massless QCD, in a Hamiltonian formulation, at zero temperature,
defined on $S^3$, where the scale of the $S^3$ is $s$. It is easy to see that, as $s\to 0$, 
$\langle \bar\psi (x) \psi (x) \rangle \sim\frac{\rm const}{s^3}$ 
with a nonzero constant. The much harder question is what happens as $s\to\infty$: perturbatively
the condensate vanishes, but, if there is S$\chi$SB, it should drop to a nonzero 
constant and level off. Now, make the temperature finite. This causes the condensate to disappear,
no matter how small the non-zero temperature is. However, if we first go to the planar limit,
the condensate will not disappear at finite and small temperatures. 

Our investigation has led us to a picture of spontaneous breaking of chiral symmetry
at infinite $N_c$ which makes the phenomenon appear generic. In other words, one almost
ends up with the conclusion that the more difficult problems are to explain how, in
more complicated cases, chiral symmetry fails to break spontaneously in the planar limit.

\section {Acknowledgments.}
R. N. acknowledges partial support by the NSF under
grant number PHY-0300065 and also partial support from Jefferson 
Lab. The Thomas Jefferson National Accelerator Facility
(Jefferson Lab) is operated by the Southeastern Universities Research
Association (SURA) under DOE contract DE-AC05-84ER40150.
H. N. acknowledges partial support 
by the DOE under grant number 
DE-FG02-01ER41165. The four dimensional results were obtained using a 48 node
cluster at Rutgers. The two dimensional results were obtained using
the 128 node cluster at Jefferson Lab as part of the LHPC collaboration.

\begin{thebibliography}{99}
\bibitem{thooft} G. 't Hooft, Nucl. Phys. B117 (1976) 519.
\bibitem{cek3} R. Narayanan, H. Neuberger, hep-lat/0303023, 
Phys. Rev. Lett. 91 (2003) 081601
\bibitem{cek4} J. Kiskis, R. Narayanan,
H. Neuberger, Phys. Lett. B574 (2003) 65.
\bibitem{gk} D. J. Gross,  Y. Kitazawa, Nucl. Phys. B206 (1982) 440.
\bibitem{2d}  J. Kiskis, R. Narayanan, H. Neuberger, Phys. Rev. D66 (2002) 025019.
\bibitem{3dphi} H. Neuberger, Phys. Rev. Lett. 60 (1988) 889; H. Neuberger,
Nucl. Phys. B300 (1988) 180.
\bibitem{overlap} H. Neuberger, Phys. Lett. B417 (1998) 141; 
H. Neuberger, Phys. Lett. B427 (1998) 353.
\bibitem{overprop} H. Neuberger, Phys. Rev. D57 (1998) 5417; H. Neuberger, 
Nuclear Physics B (Proc. Suppl.) 73 (1999) 697, hep-lat/9807009v2.
\bibitem{EHN} 
R.G. Edwards, U.M. Heller, R. Narayanan, Phys. Rev. D59 (1999) 094510.
\bibitem{quenchdiv} A. Morel, J. Phys. (FRANCE) 48 (1987) 1111.
C. W. Bernard and M. Golterman, Phys. Rev. D46 (1992) 853.
\bibitem{sharpe}
S. Sharpe, Phys. Rev. D46 (1992) 3146;
P. Damgaard, Nucl. Phys. B608 (2001) 162.
\bibitem{vw} J.J.M. Verbaarschot and T. Wettig, Ann. Rev. Nucl. Part. Sci. 50 (2000) 343.
\bibitem{sv} E.V. Shuryak and J.J.M. Verbaarschot, Nucl. Phys. A560 (1993) 306.
\bibitem{chlan} H. Leutwyler and A. Smilga, Phys. Rev. D46 (1992) 5607,
A. Smilga and J.J.M. Verbaarschot, Phys. Rev. D51 (1995) 829.
\bibitem{damsplit} P. Damgaard, K. Splittorf,  Phys. Rev. D62 (2000) 054509.
\bibitem{quenchcond} J.C. Osborn, D. Toublan and J.J.M. Verbaarschot,
Nucl. Phys. B540 (1999) 317; P. Damgaard, J.C. Osborn, D. Toublan
and J.J.M. Verbaarschot, Nucl. Phys. B547 (1999) 305.
\bibitem{dn} P.H. Damgaard and S.M. Nishigaki, hep-th/0006111 v2, 
Phys. Rev. D63 (2001) 045012.
\bibitem{akedam} G. Akemann, P H. Damgaard, Phys. Lett. B583 (2004) 199.
\bibitem{bc} T. Banks and A. Casher, Nucl. Phys. B169 (1980) 103.
\bibitem{overtop} R. Narayanan and H. Neuberger, Phys. Rev. Lee. 71 (1993) 3251;
R. Narayanan and H. Neuberger, Nucl. Phys. B443 (1995) 305.
\bibitem{single} H. Neuberger, Phys. Rev. Lett. 81 (1998) 4060.
\bibitem{double} H. Neuberger, Int. J. Mod. Phys. C10 (1999) 1051.
\bibitem{twchiu} Ting-Wai Chiu and Tung-Han Hsieh, Phys. Rev. E68 (2003) 066704.
\bibitem{zol} J. van den Eshof, A. Frommer, T. Lippert, K. Schilling
and H.A. van der Vorst, Comput. Phys. Commun. 146 (2002) 203;
T-W. Chiu, T-H. Hsieh, C-H. Huang and T-R. Huang, Phys. Rev. D66 (2002) 114502.
\bibitem{ritz} 
T. Kalkreuter and H. Simma, Comput. Phys. Commun. 93 (1996) 33.
\bibitem{teper} B. Lucini, M. Teper and U. Wenger, JHEP 0401:061 (2004).
\bibitem{harris} C. Alexandrou, H. Panagopoulos, E. Vicari, Nucl. Phys. B571 (2000) 257.
\bibitem{tadpole} P. Lepage, Lectures at 1996 Schladming Winter School on
Perturbative abd Nonperturbative Aspects of Quantum Field Theory,
Schladming, Austria, March 1996, hep-lat/9607076.
\bibitem{horsley} R. Horsley, H. Perlt, P. E. L. Rakow, G. Schierholz, A. Schiller, 
hep-lat/0404007. 
\bibitem{beci} D. Be\'cirevi\'c V. Lubicz, hep-ph/0403044. 
\bibitem{hooft2d} G. 't Hooft, Nucl. Phys. B75 (1974) 461;
G. 't Hooft, in New Phenomena in Subnuclear Physics, Part A,
Proceedings of the International School of Subnuclear Physics, Erice, 
1975 edited by A. Zichichi (Plenum, New York, 1977), Vol. 1;  
M. Burkardt, F. Lenz, M. Thies, Phys. Rev. D65 (2002) 125002;
A.R. Zhitnitsky, Phys. Lett. B165 (1985) 405; Sov. J. Nucl. Phys. 43 (1986) 999;
Yad. Fiz. 43 (1986) 1553.
\end{thebibliography}
\end{document}